%% file: main.tex
\documentclass[10pt,twocolumn,twoside]{IEEEtran} 

\IEEEoverridecommandlockouts                              

\usepackage{graphicx,amsmath,amssymb}
\usepackage[noadjust]{cite}
\usepackage{comment,color}
\usepackage{kotex}
\usepackage{algorithm}
\usepackage{algpseudocode}
\usepackage{tikz}
\usepackage{pgfplots}
\usepackage{grffile}
\pgfplotsset{compat=newest}
\usetikzlibrary{plotmarks}
\usetikzlibrary{arrows.meta}
\usepgfplotslibrary{patchplots}
\usepackage{siunitx}
\usetikzlibrary{shapes,arrows}
\tikzstyle{block} = [draw, fill=white, rectangle, minimum height=2em, minimum width=3em]
\tikzstyle{sum} = [draw, fill=white, circle, node distance=1cm]
\usetikzlibrary{decorations.pathmorphing}
\usetikzlibrary{calc}
\usetikzlibrary{arrows.meta,positioning}
\usetikzlibrary{calc}
\tikzset{
  block/.style = {draw, fill=white, rectangle, minimum height=2em, minimum width=4em, align=center},
  line/.style  = {-{Stealth[length=2.2mm]}, thick},
  plain/.style = {thick}
}
\usetikzlibrary{fit,backgrounds}

\usepackage{enumitem}
\usepackage{amsthm}

\newcommand{\ra}{\rightarrow}

\newcommand{\bbR}{\ensuremath{{\mathbb R}}}
\newcommand{\bbZ}{\ensuremath{{\mathbb Z}}}
\newcommand{\bbN}{\ensuremath{{\mathbb N}}}

\newcommand{\calD}{\mathcal{D}}

\newcommand{\calF}{\mathcal{F}}

\newcommand{\sfs}{\mathsf{s}}
\newcommand{\sfL}{\mathsf{L}}
\newcommand{\sfQ}{\mathsf{Q}}
\newcommand{\sfM}{\mathsf{M}}

\newcommand{\bfx}{\mathbf{x}}
\newcommand{\bfr}{\mathbf{r}}
\newcommand{\bfc}{\mathbf{c}}
\newcommand{\bfy}{\mathbf{y}}
\newcommand{\bfu}{\mathbf{u}}

\newcommand{\bfzero}{\mathbf{0}}

\newcommand{\rmF}{\mathrm{F}}
\newcommand{\rmG}{\mathrm{G}}

\newcommand{\rmH}{\mathrm{H}}
\newcommand{\rmJ}{\mathrm{J}}
\newcommand{\rmP}{\mathrm{P}}

\newcommand{\Enc}{\mathsf{Enc}}
\newcommand{\Dec}{\mathsf{Dec}}

\newcommand{\sk}{\mathsf{sk}}
\newcommand{\nom}{\mathsf{nom}}

\newcommand{\ini}{\mathsf{ini}}
\newcommand{\Adv}{\mathsf{Adv}}
\newcommand{\View}{\mathsf{View}}

\newcommand{\sample}{\mathrel{\leftarrow_{\$}}}

\newcommand{\llceil}{\left\lceil}

\newcommand{\rrfloor}{\right\rfloor}
\newcommand{\modp}{~\mathrm{mod}~}

\newtheorem{thm1}{\bf Theorem}
\newtheorem{prop1}{\bf Proposition}
\newtheorem{lem1}{\bf Lemma}
\newtheorem{assmpt1}{\bf Assumption}
\newtheorem{defn1}{\bf Definition}
\newtheorem{rem1}{\bf Remark}

\newtheorem{cor1}{\bf Corollary}

\newtheorem{prob1}{\bf Problem}

\newenvironment{asm}{\begin{assmpt1}}{\hfill$\square$\end{assmpt1}}
\newenvironment{rem}{\begin{rem1}}{\hfill$\square$\end{rem1}}
\newenvironment{lem}{\begin{lem1}}{\hfill$\square$\end{lem1}}
\newenvironment{thm}{\begin{thm1}}{\hfill$\square$\end{thm1}}

\newenvironment{prop}{\begin{prop1}}{\hfill$\square$\end{prop1}}
\newenvironment{prob}{\begin{prob1}}{\hfill$\square$\end{prob1}}

\title{\LARGE \bf
A Learning With Errors based Encryption Scheme for Dynamic Controllers that Discloses Residue Signal for Anomaly Detection
}

\author{Yeongjun Jang, Joowon Lee, Junsoo Kim, Takashi Tanaka, and Hyungbo Shim
\thanks{*This work was supported by Institute of Information \& communications Technology Planning \& Evaluation (IITP) grant funded by the Korea government(MSIT) (No.RS-2024-00441762, Global Advanced Cybersecurity Human Resources Development)
}
\thanks{Y.~Jang and H.~Shim are with ASRI, Department of Electrical and Computer Engineering, Seoul National University, Seoul, 08826, Korea (email: jangyj0512@snu.ac.kr, hshim@snu.ac.kr).
}
\thanks{J.~Lee is with Department of Decision and Control Systems, School of Electrical Engineering and Computer Science, KTH Royal Institute of Technology, Stockholm, 114 28, Sweden (email: joowon@kth.se).
}
\thanks{T.~Tanaka is with the School of Aeronautics and Astronautics and the Elmore Family School of Electrical and Computer Engineering, Purdue University, West Lafayette, IN 47907, USA (email: tanaka16@purdue.edu).
}
\thanks{J.~Kim is with the Department of Electrical and Information Engineering,
Seoul National University of Science and Technology, Seoul, 01811, Korea (email: junsookim@seoultech.ac.kr).
}
}

\begin{document}

\maketitle
\thispagestyle{plain} 
\pagestyle{plain} 


\begin{abstract}
    Although encrypted control systems ensure confidentiality of private data, it is challenging to detect anomalies without the secret key as all signals remain encrypted. To address this issue, we propose a homomorphic encryption scheme for dynamic controllers that automatically discloses the residue signal for anomaly detection, while keeping all other signals private. To this end, we characterize the zero-dynamics of an encrypted dynamic system defined over a finite field of integers and incorporate it into a Learning With Errors (LWE) based encryption scheme. We then present a method to further utilize the disclosed residue signal for implementing dynamic controllers over encrypted data, without requiring re-encryption even when they have non-integer state matrices.
\end{abstract}

\section{Introduction}
The threat of cyber-attacks against networked control systems has increased due to the development of attack strategies that exploit vulnerabilities in communication channels and computing devices to avoid detection \cite{Lang11,SandAmin15,TeixSham15,ParkShim16}.
To address such privacy concerns, the notion of \textit{encrypted control} \cite{KogiFuji15,KimjLeec16,DaruAlex21,KimjKimd22,SchlBinf23} has emerged, which ensures the confidentiality of all data during transmission and computation stages by integrating homomorphic encryption (HE) into control systems---a cryptographic technique that enables direct evaluation of arithmetic operations on encrypted data without requiring decryption.
Recent related studies have mainly focused on concealing control parameters and signals while preserving control performance, with applications across various domains, including model predictive control \cite{AlexMora18,KadaAbdu24}, average consensus \cite{Kish18,Hadj20,LeedKimj20,TeraKogi25}, and reinforcement learning \cite{DzurVala24,SuhjJang25}.

While ensuring confidentiality is crucial, it is also important to develop defense strategies that enable anomaly and fault detection, thereby safeguarding the system against data corruption attacks.
However, a fundamental challenge is that encryption, which protects data confidentiality, also obscures the information required for attack detection.
For example, consider a network-side anomaly detector operating within an encrypted control system, as depicted in Fig.~\ref{fig:config}. 
Its objective is to trigger an alarm when the residue signal, which represents deviation from the expected behavior, exceeds a prescribed threshold. 
This approach has been shown to be effective against a broad class of attacks, including false data injection (FDI) and denial-of-service (DoS) attacks \cite{SandGupt22}.
Yet, because the residue remains encrypted, performing this operation would require access to the secret key, and sharing the secret key with a network-side detector may introduce security risks.

To address this limitation, \cite{MartZhan19} suggested transmitting the encrypted residue to a trusted entity that holds the secret key, and delegating detection.
Alternatively, \cite{AlexBurb22} proposed comparing the residue with the threshold directly in the encrypted domain, which can be computationally intensive. 
Moreover, the comparison result remains encrypted, thus still requiring decryption by an entity holding the secret key. 
Consequently, these approaches introduce additional communication and/or computation overhead, which may be critical in real-time operations. 
Fundamentally, the existence of anomalies in encrypted control systems has not been identifiable without the secret key.

\begin{figure}[t]
\centering
\begin{tikzpicture}
\node[block, name = actuator, line width=1pt,scale=0.8]{$\mathsf{Decryption}$};
\node[block, right of =actuator, node distance = 3.5cm, name = controller, fill=black!15, fill opacity=1, line width=1pt,scale=0.8]{ $\begin{aligned}
        \bfx(t+1) &= \rmF\cdot\bfx(t) + \rmG\cdot\bfy(t) \\
        \bfu(t) &= {\rmP} \cdot \bfx(t) 
  \end{aligned}$ };
\node[block, right of = controller, node distance = 3.5cm, name = sensor, line width=1pt,scale=0.8]{$\mathsf{Encryption}$};
\node[block, below of = controller, node distance = 1.8cm, name = residue,fill=black!15, fill opacity=1,line width=1pt, scale=0.8]{$\mathsf{Alarm}(\bfr(t),\mathsf{Threshold})$};
\node[above of = actuator, node distance = 0.7cm, name = actuatorname, scale=0.8]{\textit{Actuator}};
\node[above of = sensor, node distance = 0.7cm, name = sensorname, scale=0.8]{\textit{Sensor}};
\node[above of = controller, node distance = 0.7cm, name = controllername, scale=0.8]{\textit{Controller (over encrypted data)}};
\node[below of = residue, node distance = 0.7cm, name = residuename, scale=0.8]{\textit{Anomaly detector (over encrypted data)}};
\draw[line width=1pt, ->] (controller) -- (actuator)node[pos = 0.5, above,scale=0.8]{  $\bfu(t)$}; 
\draw[line width=1pt,->] (sensor) -- (controller)node[pos = 0.5, above,scale=0.8]{ $\bfy(t)$};
\draw[line width=1pt,->] (controller) -- (residue)node[pos = 0.5, right,scale=0.8]{\begin{tabular}{l}
      $\bfr(t) = {\rmH}\cdot\bfx(t) + {\rmJ}\cdot\bfy(t)$ \\
      (encrypted residue)
    \end{tabular}};
\end{tikzpicture}
\caption{Configuration of encrypted control system with anomaly detector. } \label{fig:config}
\end{figure}

In this paper, we propose a homomorphic encryption scheme for dynamic controllers that automatically discloses the residue signal, while keeping all other signals private.
Importantly, the disclosure is achieved without decryption, even though the encrypted residue is computed directly from encrypted states and inputs.
Our key idea is to enforce the \textit{masking term} of the encrypted residue to remain identically zero, thereby enabling automatic disclosure.
Specifically, we build upon the Learning With Errors (LWE) based encryption scheme of \cite{Rege09}, which ensures confidentiality by adding a masking term to the message.
By leveraging the notion of zero-dynamics \cite{Khal96}, we appropriately manipulate the encrypted initial state and inputs, so that the masking term of the encrypted residue remains identically zero.
As a result, a network-side detector can directly detect anomalies without requiring access to the secret key or performing additional computations in the encrypted domain.
We provide a security analysis showing that the proposed method does not compromise the security of the standard LWE based scheme beyond disclosing the residue signal.

Furthermore, we present a method to utilize the disclosed residue signal for implementing dynamic controllers over encrypted data, without requiring re-encryption.
Re-encryption, proposed in \cite{KimjShim23} and also used in \cite{TeraSada23,LeejLeed24}, refers to feeding the output of an encrypted controller, which is decrypted and encrypted at the actuator, back to the controller as an input.
This procedure is typically used to convert the state matrix of the controller into an integer matrix, a property required for encrypted dynamic controllers \cite{CheoHank18}.
However, it necessitates an additional communication link and imposes extra computational burden on the actuator.
Instead, since the residue signal is already disclosed in our scheme, we use it directly as the fed-back input to the controller, without transmitting it to the actuator or performing decryption (See Fig.~\ref{fig:diagram}).
We provide a guideline for choosing the encryption parameters under which the resulting encrypted controller preserves the performance of the unencrypted controller.

The remainder of this paper is organized as follows. Section~\ref{sec:problem} reviews the standard LWE based HE scheme and formulates the problem. 
Section~\ref{sec:main} presents the proposed encryption scheme, and analyzes its correctness and security properties. Section~\ref{sec:application} applies the scheme to dynamic controllers and establishes a condition for the parameters, under which a desired performance is guaranteed. 
Section~\ref{sec:simul} presents simulation results. 
Section~\ref{sec:conclusion} concludes the paper.

\textit{Notation:} 
Let $\bbR$, $\bbZ$, and $\bbZ_{\ge 0}$ denote the sets of real numbers, integers, and non-negative integers, respectively.
The floor and rounding operations are denoted by $\lfloor \cdot \rfloor$ and $\lceil \cdot \rfloor$, respectively.
For $q\in\bbN$, we define the set 
$\bbZ_q:=\bbZ\cap[-q/2,q/2)$
and the modulo operation by $a\modp q := a- \lfloor (a+q/2)/q \rfloor q$ for all $a\in\bbZ$.
The floor, rounding, and modulo operations are defined component-wisely for vectors and matrices. 
For a sequence $v_1,\ldots, v_n$ of scalars or matrices, we define $[v_1;\cdots;v_n]:=[v_1^\top,\ldots,v_n^\top]^\top$.
The zero matrix and the identity matrix are denoted by $\bfzero_{m\times n}\in\bbZ^{m\times n}$ and $I_n\in\bbZ^{n\times n}$, respectively.
For vectors and matrices, $\|\cdot\|$ denotes the (induced) infinity norm. 
For a finite set $S$, we use $s\sample S$ to indicate that $s$ is sampled uniformly at random from $S$.

\section{Preliminaries and Problem Formulation}\label{sec:problem}

\subsection{LWE based encryption scheme}\label{subsec:LWE}

The Learning With Errors (LWE) based encryption scheme of \cite{Rege09} is introduced, focusing on its additively homomorphic property.
We consider the set $\bbZ_q$ with the modulus $q\in \bbN$ as the space of plaintexts (messages to be encrypted). 
Let the secret key $\sk\in\bbZ_q^N$ of length $N\in\bbN$ be sampled from the set of ternary vectors $\{-1,0,1\}^N$.
Given an $h$-dimensional \textit{plaintext} $m\in\bbZ_q^h$, encryption is performed as 
\begin{equation}\label{eq:standardLWE}
        \Enc(m) := \begin{bmatrix}
        	 m + b, &\!\! A
        \end{bmatrix}\modp q\in\bbZ_q^{{h\times(N+1)}},
\end{equation}
where $A\sample\bbZ_q^{h\times N}$ is a \textit{random matrix term} and the \textit{masking term} $b\in\bbZ_q^h$ is computed as 
$$b:=A\cdot \sk + e$$ 
with an \textit{error term} $e\in\bbZ^h$. 
Each element of $e$ is independently drawn from the zero-mean discrete Gaussian distribution, denoted by $\calD_\sigma^\delta$, which has standard deviation $\sigma\ge0$ and is truncated to $[-\delta,\delta]$ for some $\delta>0$. 
This essentially makes $m+b\modp q$ appear random in $\bbZ_q^h$, thereby hiding the plaintext $m$.

Given the secret key $\sk$, a ciphertext (encrypted message) $\bfc\in\bbZ_q^{{h\times(N+1)}}$ can be decrypted as
\begin{equation*}
    \Dec(\bfc):= \bfc \begin{bmatrix}
	1\\-\sk
    \end{bmatrix} \modp q\in\bbZ_q^h.
\end{equation*}
This allows the plaintext $m$ to be recovered from $\Enc(m)$ as
\begin{equation*}
\begin{aligned}
	\Dec(\Enc(m))
	&=\begin{bmatrix}
		 m + b, &\!\! A
	\end{bmatrix} \begin{bmatrix}
		1\\-\sk
	\end{bmatrix} \modp q \\
	&=  m + e\modp q,
\end{aligned}
\end{equation*}
along with the error term $e$.
Throughout the paper, we omit the operation $\mathrm{mod}~q$ in the arguments of encryption and decryption algorithms for simplicity.

It follows from the definition of $\Dec$ that the described scheme is additively homomorphic, that is, 
\begin{equation*}
    \Dec(\bfc_1+\bfc_2) = \Dec(\bfc_1) + \Dec(\bfc_2) \modp q, 
\end{equation*}
for all $\bfc_1 \in \bbZ_q^{h\times(N+1)}$ and $\bfc_2 \in \bbZ_q^{h\times(N+1)}$.
Therefore, a matrix $K\in\bbZ^{l\times h}$ can be multiplied to $\Enc(m)$ in \eqref{eq:standardLWE}, as
\begin{align}\label{eq:multCiph}
    &K\cdot \Enc(m) \modp q \nonumber \\
    &=\begin{bmatrix}
        	 Km + Kb, &\!\! KA
        \end{bmatrix}\modp q \in\bbZ_q^{{l\times(N+1)}},
\end{align}
yielding an $l$-dimensional ciphertext, which is decrypted as
\begin{equation*}
	\Dec(K\cdot \Enc(m))
	= K( m+e)\modp q \in\bbZ_q^l.
\end{equation*}
Note that the multiplication by the matrix $K$ is applied to both the plaintext $m$ and the error term $e$.
We refer to $Km$, $KA$, $Kb$, and $Ke$ as the plaintext, random matrix term, masking term, and error term, respectively, of the ciphertext $K\cdot \Enc(m)$.

The security of the described scheme relies on the computational hardness of the LWE problem.
Roughly, it is \textit{hard} to distinguish samples of the form \eqref{eq:standardLWE} from the same number of samples drawn uniformly at random from $\bbZ_q^{h\times(N+1)}$, which underlies the standard notion of semantic security \cite{Rege09}; see \cite{Rege09,AlbrPlay15,Lind17} for comprehensive discussions.
The level of security depends on the choice of the parameters $(N,q,\sigma)$. 
A guideline for selecting suitable parameters that achieve a desired security level is provided in \cite{Rege09}. 
Conversely, \cite{AlbrPlay15} provides a practical tool called the LWE estimator that estimates the security level achievable by a given set of parameters $(N, q, \sigma)$.

\begin{rem}\label{rem:L}\upshape
    In practice, the effect of the error term injected during encryption can be negated by scaling the plaintext by a sufficiently large number.
    For example, consider a plaintext $m\in\bbZ_q$ and a scale factor $1/\sfL\in\bbN$ that satisfy $\sfL<1/2\delta$ and $|m|<\sfL\cdot( q/2-\delta)$.
    Then, $|m/\sfL + e| < q/2$ for any $e\in\bbZ$ drawn from $\calD_\sigma^\delta$.
    Thus, it holds that 
    \begin{equation}\label{eq:errEffect}
        \left\lceil\sfL\cdot\Dec(\Enc(m/\sfL))\right\rfloor =  \left\lceil\sfL\cdot \left(\frac{m}{\sfL} + e\modp q\right)\right\rfloor  = m ,
    \end{equation}
    since $a=a\modp q$ for all $a\in\bbZ$ such that $|a|<q/2$, and $|\sfL\cdot e|<1/2$.
\end{rem}

\subsection{Problem formulation}

Consider a discrete-time single-input single-output dynamic controller that operates over the plaintext space $\bbZ_q$ with the modular arithmetic, written by
\begin{align}\label{eq:sys}
    x_q(t+1) &= Fx_q(t) + Gy_q(t) \modp q, ~~~~ x_q(0) = x_{q}^\ini , \nonumber\\
    u_q(t) &= Px_q(t)\modp q, \\
    r_q(t) &= Hx_q(t) + Jy_q(t)\modp q, \nonumber
\end{align}
where $x_q(t)\in\bbZ_q^n$ is the state with the initial value $x_{q}^\ini\in\bbZ_q^n$, $y_q(t)\in\bbZ_q$ is the input, $u_q(t)\in\bbZ_q$ is the output, and $r_q(t)\in\bbZ_q$ is the residue signal for anomaly detection.
The design of the residue signal $r_q(t)$ is not of interest in this paper and is assumed to be given.
We also assume that the control parameters are publicly known and are given as 
\begin{equation}\label{eq:sysParams}
    F\in\bbZ^{n\times n}, ~  G\in\bbZ^{n}, ~  P\in\bbZ^{1\times n}, ~  H\in\bbZ^{1\times n}, ~  J\in\bbZ,
\end{equation}
which ensures that $\{x_q(t),u_q(t),r_q(t)\}$ retain their values in $\bbZ_q$ under the modular arithmetic of \eqref{eq:sys}.

Dynamic controllers in real-world applications usually operate over $\bbR$ and are not limited to $\bbZ_q$. 
However, in order to implement them using the LWE based scheme, it is necessary to first convert the controllers to operate over the plaintext space $\bbZ_q$. 
Therefore, we assume that such conversion has been completed a priori, and develop our discussions based on this premise.
A method to perform such conversion will be discussed in Section~\ref{sec:application} with detail.

\begin{rem}\label{rem:integer}\upshape
    Various strategies to convert a dynamic controller to operate over $\bbZ_q$ have been investigated in the literature.
    A possible strategy is to first transform the state matrix $F$ of a given controller into an integer matrix while preserving the same input-output relation, following the methods of, for example, \cite{KimjShim21,Tava22,KimjShim23}.
    Alternatively, one may directly design a stabilizing controller having an integer state matrix, as in \cite{LeejLeed25}.
    Subsequently, all signals and parameters other than the (integer) state matrix are scaled by a sufficiently large scale factor and then rounded to integers \cite{KimjKimd22,SchlBinf23}. 
    The resulting \textit{integerized} system is then projected onto $\bbZ_q$ by taking the modulo operation.
\end{rem}

Now, we describe the configuration of the encrypted controller depicted in Fig.~\ref{fig:config}, and specify the problem of interest.
Let the controller \eqref{eq:sys} over $\bbZ_q$ operate based on the introduced LWE based scheme, as
\begin{align}\label{eq:sysEncEx}
    \bfx(t+1) &= F \cdot \bfx(t) + G \cdot \Enc(y_q(t)) \modp q, \nonumber\\
        \bfu(t) &= P\cdot \bfx(t) \modp q, \nonumber \\
        \bfr(t) &= H\cdot \bfx(t) + J \cdot \Enc(y_q(t)) \modp q, \\
        \bfx(0) &= \Enc(x_{q}^\ini),\nonumber
\end{align}
where $\bfx(t)\in\bbZ_q^{n\times(N+1)}$, $\bfu(t)\in\bbZ_q^{1\times(N+1)}$, and $\bfr(t)\in\bbZ_q^{1\times(N+1)}$ are the state, the output, and the residue signal as ciphertexts, respectively.

We model the adversary as either an eavesdropper or the encrypted controller \eqref{eq:sysEncEx} itself, who collects the ciphertexts $\Enc(x_{q}^\ini)$ and $\{\Enc(y_q(\tau))\}_{\tau=0}^\infty$ (with which $\{\bfx(\tau),\bfu(\tau),\bfr(\tau)\}_{\tau=0}^\infty$ can also be obtained). 
The adversary's goal is to infer any meaningful information about the underlying plaintexts.
Thanks to the security of the LWE based scheme, the collected ciphertexts essentially reveal no information to the adversary.

However, a problem in this existing setup is that it also becomes difficult for a network-side detector to monitor and detect anomalies.
Based on this motivation, \textit{we suggest that the encryption scheme be modified, so that the residue signal is automatically disclosed as a plaintext.}
This will enable the network-side detector to directly detect anomalies without requiring access to the secret key.
At the same time, for security, the modified scheme should not leak additional information beyond what is implied by the disclosed residue signal, which will be formalized in Section~\ref{subsec:scheme}.
The problem is more specifically stated as follows.

\begin{prob}\label{prob}\upshape
    Given the parameters $\{F,G,P,H,J,x_q^\ini\}$ of \eqref{eq:sys}, modify the encryption algorithm $\Enc$ in \eqref{eq:standardLWE}, so that the encrypted controller \eqref{eq:sysEncEx} automatically discloses the plaintext $r_q(t)$ of $\bfr(t)$ without decryption.
    The modification should not compromise the security of the standard LWE based scheme beyond disclosing the residue signal.
\end{prob}

\section{Proposed Encryption Scheme}\label{sec:main}
This section serves to describe the proposed encryption scheme.
Let us rewrite the ciphertexts $\Enc(x_{q}^\ini)$ and $\Enc(y_q(t))$ of \eqref{eq:sysEncEx} in the form of \eqref{eq:standardLWE}, as
\begin{subequations}\label{eq:standardEncEx}	
     \begin{align}
		\Enc(x_{q}^\ini) &= \begin{bmatrix}
			x_{q}^\ini+ b_{x}^\ini,& \!\! A_x^\ini
		\end{bmatrix} \modp q,\\
		\Enc(y_q(t)) &= \begin{bmatrix}
			y_q(t)+b_y(t),&\!\! A_y(t)
		\end{bmatrix} \modp q, \notag
	\end{align}
where $A_x^\ini\sample\bbZ_q^{n\times N}$ and $A_y(t) \sample \bbZ_q^{1\times N}$ for each $t\ge 0$. 
The masking terms are given by
\begin{align}
	\begin{split}
	b_{x}^\ini&:= A_x^\ini\cdot \sk + e_x^\ini\modp q\in\bbZ_q^n,\\
    b_y(t)&:= A_y(t)\cdot \sk + e_y(t)\modp q\in\bbZ_q,
\end{split}\label{eq:standardEncExB}
\end{align}
\end{subequations}
where each element of the error terms $e_x^\ini\in\bbZ^{n}$ and $e_y(t)\in\bbZ$ (for each $t\ge 0$) is drawn from the distribution $\calD_\sigma^\delta$.

Thanks to the linearity of \eqref{eq:sysEncEx}, the encrypted residue signal $\bfr(t)$ can be written as
\begin{equation}\label{eq:encRes}
    \bfr(t) = 
    \begin{bmatrix}
        	 r_q(t) + b_r(t), &\!\! A_r(t)
        \end{bmatrix}\modp q,
\end{equation}
where the plaintext $r_q(t)$ is identical to the residue signal of \eqref{eq:sys}, and the masking term $b_r(t)\in \bbZ_q$ evolves through the following dynamics\footnote{The random matrix term $A_r(t)$ also obeys the same dynamics as \eqref{eq:bDyn}, which enables the decryption of $\bfr(t)$ using the secret key $\sk$.} over $\bbZ_q$:
\begin{equation}\label{eq:bDyn}
    \begin{split}
        b_x(t+1) &= Fb_x(t) + Gb_y(t)\modp q, \quad b_x(0) = b_x^\ini, \\
        b_r(t) &= Hb_x(t) + Jb_y(t)\modp q.
    \end{split}
\end{equation}
Our goal is to modify the encryption algorithm $\Enc$, so that 
\begin{equation*}
    b_r(t) \equiv 0, \quad \forall t\ge 0,
\end{equation*}
thus disclosing $r_q(t)$ in \eqref{eq:encRes} without decrypting $\bfr(t)$.

Observe that $b_r(t)$ can be considered as the output of the system \eqref{eq:bDyn}.
In this context, we aim to figure out a condition on the initial state $b_x^\ini$ and the input sequence $\{b_y(\tau)\}_{\tau=0}^\infty$ under which $b_r(t)$ remains identically zero. 
As a first step, we investigate the \emph{zero-dynamics} of dynamic systems \textit{over the space $\bbZ_q$.}

\subsection{Zero-dynamics of systems over $\bbZ_q$}\label{subsec:zeroDyn}

We follow the procedure of \cite[Chapter~13]{Khal96} to derive the Byrnes-Isidori normal form of the system \eqref{eq:bDyn}.
Before proceeding, let us fix the modulus $q$ as a prime number, so that $\bbZ_q$ equipped with the modular addition and multiplication becomes a field \cite[Chapter~2.3]{Hung12}.
Then, for any nonzero element $a\in\bbZ_q$, there exists a unique multiplicative inverse $a^{-1}\in\bbZ_q$ such that $aa^{-1}~\mathrm{mod}\, q = 1$.
This allows us to regard $\bbZ_q^n$ as a \emph{vector space over the field $\bbZ_q$}, enabling the use of standard linear algebraic notions---such as linear independence, basis, and rank---that have been developed for arbitrary fields in \cite[Chapter~1]{Frie14}.

We define the relative degree $\nu\ge 0$ of the system \eqref{eq:bDyn} as
\begin{align*}
    \nu := \begin{cases}
	0, &\text{if}~J\modp q\neq 0,\\
	d, &\text{if}~J\modp q=0,~HF^{d-1}G\modp q\neq 0, \\
    &\text{and}~HF^iG\modp q = 0,~~\forall i\in\{0,\ldots, d-2\}, 
\end{cases}
\end{align*}
analogous to that of linear systems over $\bbR$.
We defer the discussion for the case $\nu=0$ and obtain the normal form for the case $\nu \ge 1$ first.

By virtue of linear algebra over the field $\bbZ_q$, the row vectors in $\{HF^{i}\modp q\mid i=0,\ldots,\nu-1\}\subset \bbZ_q^{1\times n}$ are linearly independent. 
Indeed, suppose that 
\begin{equation*}
    a_0 H + a_1 HF+\ldots+a_{\nu-1}HF^{\nu-1}\modp q = \bfzero_{1\times n}
\end{equation*}
for some $a_i\in\bbZ_q$ for $i=0,\ldots,\nu-1$. 
By consecutively multiplying $F^{i}G$ from the right for $i=0,1,\ldots,\nu-1$, we have $a_i=0$ for $i=\nu-1,\nu-2,\ldots,0$ from the definition of relative degree.
This allows us to construct a full row rank matrix
\begin{equation}\label{eq:T_2}
	T_2:=  \begin{bmatrix}
		H\\ HF \\\vdots\\ HF^{\nu-1}
	\end{bmatrix}\modp q\in\bbZ_q^{\nu\times n}.
\end{equation}

Since $T_2$ is of full row rank and $HF^{\nu-1}G\modp q\ne 0$, we can choose a matrix $T_1\in\bbZ_q^{(n-\nu) \times n}$ such that $[T_1;T_2] \in \bbZ_q^{n \times n}$ is invertible and 
\begin{equation}\label{eq:TG}
    \begin{bmatrix}
        T_1\\
        T_2
    \end{bmatrix}G  \modp q =
    \begin{bmatrix}
        \bfzero_{(n-1)\times 1} \\
        g
    \end{bmatrix} \in \bbZ_q^n,
\end{equation}
where 
\begin{equation}\label{eq:g}
    g := HF^{\nu-1}G \modp q\ne 0.
\end{equation}
Namely, the upper $n-1$ rows of the matrix $[T_1;T_2]$ form a basis of the space $\{w\in\bbZ_q^{1\times n} \mid wG\modp q =0\}$, whose dimension is clearly $n-1$.

Next, let $[V_1, V_2]\in\bbZ_q^{n\times n}$ with $V_1 \in \bbZ_q^{n \times (n-\nu)}$ and $V_2\in \bbZ_q^{n \times \nu}$ denote the inverse matrix of $[T_1; T_2]$:
\begin{align}\label{eq:TV}
    \begin{bmatrix}
            V_1, &\!\! V_2    
        \end{bmatrix}
        \begin{bmatrix}
        T_1\\
        T_2
        \end{bmatrix} 
        \modp q &= \begin{bmatrix}
        T_1\\
        T_2
        \end{bmatrix} 
        \begin{bmatrix}
            V_1, &\!\! V_2    
        \end{bmatrix}\modp q \\
        &= 
        \begin{bmatrix}
            T_1V_1, &\!\! T_1V_2 \\
            T_2V_1, &\!\! T_2V_2
        \end{bmatrix} \modp q = I_n. \nonumber
\end{align}

The following proposition presents the normal form representation of the system \eqref{eq:bDyn}, which is obtained by a coordinate transformation using the matrix $[T_1;T_2]$.

\begin{prop}\label{lem:zero}\upshape
    Suppose that the system \eqref{eq:bDyn} has relative degree $\nu\ge 1$. 
    Then, the coordinate transformation
    \begin{equation} \label{eq:transform}
		\begin{bmatrix}
			b_z(t)\\ b_w(t)	
		\end{bmatrix}:= 
        \begin{bmatrix}
	   T_1 \\ T_2	    
		\end{bmatrix} b_x(t)\modp q.
	\end{equation}
yields the \textit{normal form} of \eqref{eq:bDyn}, written by
\begin{align}\label{eq:normal}
	\begin{split}
	b_z(t+1)&= F_{1}b_z(t) + F_{2}b_w(t)\modp q,\\
	b_{w_1}(t+1) &= b_{w_2}(t),\\
	\vdots\\
	b_{w_{\nu-1}}(t+1) &= b_{w_\nu}(t),\\
	b_{w_\nu}(t+1) &= \psi b_z(t) + \phi b_w(t) + gb_y(t)\modp q,\\
	b_r(t) &= b_{w_1}(t),
\end{split}
\end{align}
where $b_w(t)=:[b_{w_1}(t); \cdots; b_{w_\nu}(t)]\in\bbZ_q^\nu$ and
\begin{equation*}
\begin{aligned}
	F_1&:=T_1 F V_1\modp q, & F_2&:= T_1FV_2\modp q,\\
	\psi &:=  HF^\nu V_1\modp q, & \phi &:=  HF^\nu V_2\modp q.
\end{aligned}
\end{equation*}
\end{prop}

\begin{proof}
It follows from \eqref{eq:TV} and \eqref{eq:transform} that
\begin{equation*}
    b_x(t) = 
    V_1 b_z(t)+ V_2 b_w(t)\modp q.
\end{equation*}
Hence, it is obtained from \eqref{eq:bDyn} that
\begin{align*}
    \!b_z(t+1) \!&=\!  T_1b_x(t+1) \modp q\\
    \!&=\!T_1F V_1 b_z(t) + T_1 FV_2  b_w(t)+T_1Gb_y(t) \modp q, \\
	b_w(t+1) \!&=\! T_2b_x(t+1)\modp q  \\
    \!&=\!T_2F V_1 b_z(t) + T_2 FV_2 b_w(t) + T_2Gb_y(t)\modp q ,\\
	\!b_r(t) \!&=\! Hb_x(t)\modp q \\
    \!&=\! HV_1b_z(t)+ HV_2b_w(t) \modp q. 
\end{align*}
Since $T_2V_1 \modp q= \bfzero_{\nu \times (n-\nu)}$ and $T_2V_2\modp q = I_{\nu}$, it can be derived from \eqref{eq:T_2} that
\begin{align*}
	\begin{split}
	T_2FV_1\modp q &= \begin{bmatrix}
		\bfzero_{(\nu-1)\times (n-\nu)} \\ \psi 
	\end{bmatrix}\in\bbZ_q^{\nu\times (n-\nu)},\\
        T_2FV_2\modp q &= 
        \begin{bmatrix}
	~{\begin{bmatrix}
		\bfzero_{(\nu-1)\times 1}, &\!\! I_{\nu-1}	
	\end{bmatrix}~} \\
        \phi 
        \end{bmatrix} \in \bbZ_q^{\nu\times \nu}, \\
    HV_1\modp q &= \bfzero_{1\times (n-\nu)}, \\
    HV_2\modp q&=[1,\bfzero_{1 \times (\nu-1)}] \in\bbZ_q^{1\times \nu}.
\end{split}
\end{align*}
Also, it follows from \eqref{eq:TG} that $T_1G\modp q = \bfzero_{(n-\nu)\times 1}$ and $T_2G\modp q = [\bfzero_{(\nu-1)\times 1}; g]$. 
This concludes the proof.
\end{proof}

Analogous to the standard definition for linear systems over $\bbR$, we define the \textit{zero-dynamics} of the system \eqref{eq:bDyn} as the subsystem of \eqref{eq:normal} constrained to  $b_r(t)\equiv0$, written by
\begin{align}\label{eq:zeroDynState}
    b_z'(t+1) &= F_1b_z'(t)\modp q,\\
    b_z'(0) &= T_1b_x^\ini \modp q, \nonumber
\end{align}  
where $b_z'(t)\in\bbZ_q^{n-\nu}$.

The following lemma states that we can enforce the output $b_r(t) $ to remain identically zero by canceling certain portions of the initial state $b_{x}^\ini$ and the input sequence $\{b_y(\tau)\}_{\tau=0}^\infty$.
This aspect will play a key role in designing the proposed encryption scheme.
For clarity, we often denote $b_r(t)$ by
\begin{equation}\label{eq:rexplicit}
    b_r(t;b_x^\ini,\{b_y(\tau)\}_{\tau=0}^t)
\end{equation}
to emphasize its dependence on the initial state $b_x^\ini$ and the input sequence $\{b_y(\tau)\}_{\tau=0}^t$ up to the time step $t$.

\begin{lem}\label{lem:nu1}\upshape
    Suppose that the system \eqref{eq:bDyn} has relative degree $\nu\ge 1$.
    Given $b_x^\ini\in\bbZ_q^n$ and $b_y(\cdot):\bbZ_{\ge 0}\ra \bbZ_q$, there exist $b_w'\in\bbZ_q^\nu$ and $b_y'(\cdot):\bbZ_{\ge 0}\ra \bbZ_q$ such that 
\begin{equation} \label{eq:lem1_toShow}
    \begin{split}
       b_r(t;b_x^\ini - V_2 b_w',\{b_y(\tau) - b_y'(\tau)\}_{\tau=0}^t) \equiv0,
    \end{split}
\end{equation}
     which are uniquely determined by
\begin{subequations}\label{eq:lem1_claim}
    \begin{align}
         b_w' &= T_2 b_x^\ini\modp q, \label{eq:lem1_claimIni}\\ 
        b_y'(t) &= b_y(t) +g^{-1}\psi b_z'(t)\modp q, \label{eq:lem1_claimInput}
    \end{align}
\end{subequations}
for all $t\ge 0$, where $b_z'(t)$ is the solution to \eqref{eq:zeroDynState}.
\end{lem}

\begin{proof}
    Consider the system \eqref{eq:bDyn} with the initial state $b_x^\ini - V_2 b_w'$ and input sequence $\{b_y(\tau) - b_y'(\tau)\}_{\tau=0}^\infty$.
    It can be observed from the normal form \eqref{eq:normal} that \eqref{eq:lem1_toShow} holds if and only if
    \begin{subequations}
    \begin{align}\label{eq:lem1iff1}
        b_w(0) &= \bfzero_{\nu\times 1}, \\
        b_{w_\nu}(t+1)&=0,\label{eq:lem1iff2}
    \end{align}
    \end{subequations}
    for all $t\ge 0$.
    Since $b_w(0) = T_2(b_x^\ini - V_2b_w') \modp q$ and $T_2V_2 = I_\nu$, \eqref{eq:lem1iff1} holds if and only if \eqref{eq:lem1_claimIni} holds.
    Given that $b_w(0) =\bfzero_{\nu\times 1}$, we have 
    \begin{equation*}
       b_{w_\nu}(1) = \psi b_z'(0) + g\left( b_y(0) - b_y'(0)\right) \modp q,
    \end{equation*}
    since $b_z(0)=b_z'(0)$.
    Hence, $b_{w_\nu}(1)=0$ if and only if \eqref{eq:lem1_claimInput} holds for $t=0$.
    
    Now, suppose that \eqref{eq:lem1iff1} holds and \eqref{eq:lem1iff2} holds for $t=0,\ldots,k-1$ for some $k \ge 1$, which implies that $b_w(t)=\bfzero_{\nu\times 1}$ for $t=0,\ldots,k$. 
    Then, it follows that $b_z(t) = b_z'(t)$ for $t=0,\ldots,k$, leading to
    \begin{equation*}
        b_{w_\nu}(k+1) = \psi b_z'(k) + g\left( b_y(k) - b_y'(k) \right)  \modp q.
    \end{equation*}
    Therefore, $b_{w_\nu}(k+1)=0$ if and only if \eqref{eq:lem1_claimInput} holds for $t=k$. 
    By induction, $b_y'(t)$ is uniquely determined as \eqref{eq:lem1_claimInput}, and this concludes the proof.
\end{proof}

For the case $\nu=0$, the normal form as in \eqref{eq:normal} does not exist because $b_w(t)$ of dimension $\nu$ is not well-defined.
However, a result similar to Lemma~\ref{lem:nu1} can be directly derived from \eqref{eq:bDyn}, as follows.

\begin{lem}\label{lem:nu0}\upshape
        Suppose that the system \eqref{eq:bDyn} has relative degree $\nu=0$, i.e., $J\ne 0$. 
        Given $b_x^\ini\in\bbZ_q^n$ and $b_y(\cdot):\bbZ_{\ge 0}\ra \bbZ_q$, there exists $b_y'(\cdot):\bbZ_{\ge 0}\ra \bbZ_q$ such that 
        \begin{equation}\label{eq:lem2_toShow}
		b_r(t;b_x^\ini, \{b_y(\tau) - b_y'(\tau) \}_{\tau=0}^t) \equiv 0,
	\end{equation}
        which is uniquely determined by  
        \begin{equation}\label{eq:lem2_claim}
		b_y'(t) =b_y(t) +J^{-1}H \left(F-GJ^{-1}H\right)^tb_x^\ini\modp q,
	\end{equation}
	for all $t\geq 0$.
\end{lem}

\begin{proof}
It follows from \eqref{eq:bDyn} that \eqref{eq:lem2_toShow} holds if and only if $b_y(t)-b_y'(t)=-J^{-1}Hb_x(t)\modp q$. 
Substituting this into the state dynamics of \eqref{eq:bDyn} results in
\begin{align}\label{eq:lem2_stateDyn}
     b_x(t+1) &= Fb_x(t)+G(b_y(t)-b_y'(t)) \nonumber \\
     &= \left(F-GJ^{-1}H\right)^{t+1}b_x^\ini \modp q,
\end{align}
and thus, $b_y'(t)$ is uniquely determined as in \eqref{eq:lem2_claim}. 
This concludes the proof.
\end{proof}

In fact, Lemma~\ref{lem:nu0} can be considered as a special case of Lemma~\ref{lem:nu1} with
\begin{align}\label{eq:nu0_params}
    \!\! T_1&=V_1= I_n, & T_2&=V_2= 0, & F_1&= F-GJ^{-1}H\modp q, \nonumber\\
    \!\! F_2&=0, & \psi &= H, & \phi&=0, ~~~~~~ g=J.
\end{align}
For this reason, we focus on the case $\nu\ge 1$ for the remainder of the section.

\begin{rem}\label{remark:zerodyn}\upshape
    We show that \eqref{eq:lem1_claim} can be explicitly computed given \eqref{eq:rexplicit} for all $t\ge 0$.
    This observation will be used to establish the security of the proposed encryption scheme in the following subsection.
    Since $b_w(t) = [b_r(t);\cdots;b_r(t+\nu-1)]$ is determined for all $t \ge 0$ given \eqref{eq:rexplicit}, $b_w(0)=T_2b_x^\ini\modp q=b_w'$ can be computed.
    Meanwhile, it follows from \eqref{eq:normal} that 
    \begin{equation*}
        b_z(t) = F_1^tT_1b_x^\ini + \sum_{\tau=0}^{t-1}F_1^{t-1-\tau} F_2b_w(\tau)\modp q.
    \end{equation*}
    Hence, $b_y'(t)$ in \eqref{eq:lem1_claim} can be computed from $\{b_w(\tau)\}_{\tau=0}^t$, as
  \begin{align}\label{eq:rem2_claim}
      &b_y'(t) =   g^{-1}\left( gb_y(t) + \psi F_1^t T_1 b_x^\ini\right)\modp q \\
    &= g^{-1}\left( gb_y(t) + \psi b_z(t) +\psi\left( F_1^t T_1 b_x^\ini - b_z(t)\right)\right)\modp q \nonumber\\
    &=g^{-1}\!\left(\! b_r(t\!+\!\nu) \!-\!\phi b_w(t)\!-\! \psi  \sum_{\tau=0}^{t-1}F_1^{t-1-\tau} F_2 b_w(\tau)\!\!
				\right)\modp q, \nonumber
  \end{align}
  where the last equality follows from the relation $b_r(t+\nu)=b_{w_\nu}(t+1)$ and \eqref{eq:normal}.
  For the case $\nu = 0$, let $b_w(t) = b_r(t)$ and substitute \eqref{eq:nu0_params}.
\end{rem}

\begin{algorithm}[t]
    \caption{Proposed encryption scheme}\label{alg:enc}
    \begin{algorithmic}[1]
    \renewcommand{\algorithmicrequire}{\textbf{Input:}}
    \Require $F\in\bbZ^{n\times n}$, $G\in\bbZ^{n}$, $H\in\bbZ^{1\times n}$, and $J\in\bbZ$
    \renewcommand{\algorithmicrequire}{\# \textbf{Encrypt initial value}}
    \Require
    \State Sample $A_x^\ini$ and generate $b_x^\ini$, as in \eqref{eq:standardEncExB} 
    \State Compute $b_w'$ in \eqref{eq:lem1_claimIni}
    \State $\Enc_\ini(x_q^\ini)\gets [x_{q}^\ini+b_x^\ini - V_2b_w',\,A_x^\ini,\,V_2b_w']\modp q$ 
    \renewcommand{\algorithmicrequire}{\# \textbf{Encrypt input}}
    \Require
    \State Initialize $b_z'(0)\gets T_1b_x^\ini \modp q$
    \State $t\gets 0$
    \State \label{line:enc}Sample $A_y(t)$ and generate $b_y(t)$, as in \eqref{eq:standardEncExB} 
    \State Compute $b_y'(t)$ in \eqref{eq:lem1_claimInput} 
    \State $\Enc_t(y_q(t)) \gets [y_q(t)+b_y(t)-b_y'(t),\,A_y(t),\,b_y'(t)]\modp q$
    \State $b_z'(t+1)\gets F_1b_z'(t)$ 
    \State $t\gets t+1$ and go to Line~\ref{line:enc}
    \end{algorithmic}
\end{algorithm}

\subsection{Proposed encryption scheme}\label{subsec:scheme}

We now present the proposed encryption scheme that automatically discloses the residue signal of \eqref{eq:sys} as a plaintext.
We have investigated in Section~\ref{subsec:zeroDyn} that $b_r(t)\equiv 0$ can be achieved by canceling certain portions of $b_x^\ini$ and $\{b_y(\tau)\}_{\tau=0}^\infty$ as in \eqref{eq:lem1_toShow}.
Based on this observation, we modify $\Enc(x_{q}^\ini)$ and $\Enc(y_q(t))$ given in \eqref{eq:standardEncEx}, and define $\Enc_\ini: \bbZ_q^n \ra \bbZ_q^{n\times (N+2)}$ and $\Enc_t: \bbZ_q \ra \bbZ_q^{1\times (N+2)}$ for all $t\ge 0$, by
\begin{align}\label{eq:enc_modified}
    \!\!\!\Enc_\ini(x_{q}^\ini) 
    \!&:=\! \left[x_{q}^\ini+b_x^\ini - V_2b_w',\,A_x^\ini,\,V_2b_w'\right]\modp q, \\
    \!\!\!\Enc_t(y_q(t)) \!&:=\! \left[y_q(t)+b_y(t)-b_y'(t),\,A_y(t),\,b_y'(t)\right]\modp q. \nonumber
\end{align}
The proposed scheme \eqref{eq:enc_modified} is \textit{dynamic} in the sense that $b_y'(t)$ defined in \eqref{eq:lem1_claimInput} is generated using the zero-dynamics \eqref{eq:zeroDynState}.
The entire procedure is summarized in Algorithm~\ref{alg:enc} and also illustrated in Fig.~\ref{fig:diagram}.

\begin{rem}\upshape
    Note that the proposed scheme is not applicable when $\nu=n$, i.e., when the zero-dynamics \eqref{eq:zeroDynState} does not exist.
    In this case, $b_x^\ini=V_2b_w'$ and $b_y(t)=b_y'(t)$, meaning that the modified encryptions in \eqref{eq:enc_modified} do not obscure the initial condition and the input signal.    
\end{rem}

Since ciphertexts now have one additional column compared to \eqref{eq:standardLWE}, we correspondingly modify $\Dec$ and define the decryption of a ciphertext $\bfc\in\bbZ_q^{h \times (N+2)}$, as
\begin{equation}\label{eq:dec_modified}
    \Dec'(\bfc) := \bfc
        \begin{bmatrix}
	1 \\ 
        -\sk \\
        1
\end{bmatrix} \modp q \in\bbZ_q^h.
\end{equation}
Then, it can be easily verified that 
\begin{equation}\label{eq:decEquiv}
    \begin{split}
        \Dec'( \Enc_\ini(x_{q}^\ini) )&=\Dec(\Enc(x_{q}^\ini)),\\ \Dec'(\Enc_t(y_q(t)))&=\Dec(\Enc(y_q(t))),\quad \forall t\ge 0,
    \end{split}
\end{equation}
hold for any $x_{q}^\ini$ and $y_q(t)$.
Hence, we regard $x_q^\ini$ and $y_q(t)$ as the plaintexts of the modified ciphertexts $\Enc_\ini(x_{q}^\ini)$ and $\Enc_t(y_q(t))$, respectively.
In addition, the proposed encryption scheme is additively homomorphic, as shown in the following proposition.

\begin{prop1}\label{prop:correctness}\upshape
    The following properties hold for any $\bfc_1\in\bbZ_q^{h\times (N+2)}$ and $\bfc_2\in\bbZ_q^{h\times (N+2)}$:
    \begin{enumerate}[leftmargin=*]
        \item $\Dec'(\bfc_1+\bfc_2) = \Dec'(\bfc_1) + \Dec'(\bfc_2)\modp q$.
        \item $\Dec'(K \cdot \bfc_1)=K\cdot\Dec'(\bfc_1)\modp q$ for any $K\in\bbZ_q^{l\times h}$.\hfill $\square$
    \end{enumerate}
\end{prop1}

Proposition~\ref{prop:correctness} and \eqref{eq:decEquiv} allow us to replace $\Enc$ in \eqref{eq:sysEncEx} with the proposed encryption scheme, as
\begin{align}\label{eq:sysEncEx2}
    \bfx(t+1) &= F \cdot \bfx(t) + G \cdot \Enc_t(y_q(t)) \modp q, \nonumber\\
        \bfu(t) &= P\cdot \bfx(t) \modp q, \nonumber \\
        \bfr(t) &= H\cdot \bfx(t) + J \cdot \Enc_t(y_q(t)) \modp q, \\
        \bfx(0) &= \Enc_\ini(x_{q}^\ini),\nonumber
\end{align}
where we slightly abuse notation and consider $\bfx(t)\in\bbZ_q^{n\times(N+2)}$, $\bfu(t)\in\bbZ_q^{1\times(N+2)}$, and $\bfr(t)\in \bbZ_q^{1\times(N+2)}$ as the ciphertexts of the state, output, and residue signal, respectively.
Let us denote each element of $\bfr(t)$ by
\begin{equation}\label{eq:bfr}
    \bfr(t) =: 
    \begin{bmatrix}
        \bfr_1(t), &\!\! \bfr_2(t), &\!\! \ldots, & \!\! \bfr_{N+2}(t)    
    \end{bmatrix}.
\end{equation}
The following theorem states that $\bfr_1(t)$ is identical to the plaintext $r_q(t)$ of $\bfr(t)$ for all $t\ge 0$.

\begin{thm}\label{thm:encryption}\upshape
     Suppose that the controller \eqref{eq:sys} and its encryption \eqref{eq:sysEncEx2} share the same signal $y_q(t)$ as an external input.
     Then,  \begin{equation}\label{eq:thm1ToShow}
            \bfr_1(t)=r_q(t),
        \end{equation}
        for all $t\ge 0$.
\end{thm}

\begin{proof}
    By the linearity of \eqref{eq:sysEncEx2}, we have 
    \begin{equation*}
        \bfr_1(t)=r_q(t) + b_r(t;b_x^\ini-V_2b_w', \{b_y(\tau)-b_y'(\tau)\}_{\tau=0}^t).
    \end{equation*}
    Then, it follows from Lemma~\ref{lem:nu1} that $b_r(t)\equiv 0$, and this concludes the proof.
\end{proof}

According to Theorem~\ref{thm:encryption}, the residue signal $r_q(t)$ for anomaly detection can be recovered as $\bfr_1(t)$, without decryption.
This enables a network-side detector to directly detect anomalies without requiring the secret key.

We now formally establish the security of the proposed encryption scheme.
For an adversary $\Adv$, let us define its \textit{view}, denoted by $\View_\Adv$, as the tuple of all data available to $\Adv$.
We consider two adversaries $\Adv_1$ and $\Adv_2$ with the following views:
\begin{align*}
    \View_{\Adv_1}&:=(\Enc_\ini(x_{q}^\ini),~ \{\Enc_\tau(y_q(\tau))\}_{\tau=0}^\infty), \\
    \View_{\Adv_2}&:= (\Enc(x_{q}^\ini),~\{\Enc(y_q(\tau))\}_{\tau=0}^\infty,~\{r_q(\tau)\}_{\tau=0}^\infty).
\end{align*}
That is, $\Adv_1$ observes the modified ciphertexts in \eqref{eq:sysEncEx2}, whereas $\Adv_2$ observes the standard ciphertexts in \eqref{eq:standardEncEx} and additionally receives the residue signal of \eqref{eq:sys} as a plaintext.

The following theorem states that the view of each adversary can be deterministically reconstructed from that of the other. 

\begin{thm}\label{thm:security}\upshape
    There exist deterministic algorithms $\calF_{1}$ and $\calF_{2}$ such that
    \begin{align*}
        \calF_{1}(\View_{\Adv_1}) &= \View_{\Adv_2},~~ \calF_{2}(\View_{\Adv_2}) = \View_{\Adv_1},
    \end{align*}
    for all $x_q^\ini\in\bbZ_q^n$ and $y_q(\cdot):\bbZ_{\ge 0} \to \bbZ_q$.
\end{thm}

\begin{proof}
    The algorithm $\calF_1$ can be constructed as follows. Given $\View_{\Adv_1}$, the residue signal $\{r_q(\tau)\}_{\tau=0}^\infty$ is obtained by running \eqref{eq:sysEncEx2}, since it is disclosed by Theorem~\ref{thm:encryption}.
    The standard ciphertexts $\Enc(x_{q}^\ini)$ and $\{\Enc(y_q(\tau))\}_{\tau=0}^\infty$ can be obtained by simply adding the last columns of  $\Enc_\ini(x_{q}^\ini)$ and $\{\Enc_\tau(y_q(\tau))\}_{\tau=0}^\infty$ to their respective first columns and discarding the last column (see \eqref{eq:enc_modified}).
    
    Conversely, given $\View_{\Adv_2}$, the algorithm $\calF_2$ first computes $\{b_r(\tau)\}_{\tau=0}^\infty$ by running \eqref{eq:sysEncEx} and subtracting the known $r_q(t)$ from the first column of $\bfr(t)$. 
    Then, following the procedure described in Remark~\ref{remark:zerodyn}, it computes $b_w'$ and $b_y'(t)$ in \eqref{eq:lem1_claim}, and consequently, reconstructs $\Enc_\ini(x_{q}^\ini)$ and $\{\Enc_\tau(y_q(\tau))\}_{\tau=0}^\infty$.
    This concludes the proof.
\end{proof}

Theorem~\ref{thm:security} implies that $\View_{\Adv_1}$ is \textit{perfectly indistinguishable} from $\View_{\Adv_2}$ (and vice versa) in the sense of simulation based security \cite{Gold01}.
This guarantees that $\Adv_1$ cannot acquire any information beyond what is already implied by the view of $\Adv_2$.
Therefore, the proposed encryption scheme does not compromise the security of the standard LWE based scheme beyond the intended disclosure of the residue signal.

\begin{rem}\upshape
    One may be concerned that disclosing the residue signal could reveal some sensitive information about the system.
    However, even in attack-free scenarios, residue signals are highly influenced by noise, disturbances, and model uncertainties \cite{PasqDorf13}.
    Thus, the residue signal alone tends to be noisy and generally uninformative in practice. 
    Moreover, if \eqref{eq:controller} is constructed as an observer based controller, as in Section~\ref{sec:simul}, the residue signal may reveal some information about the state estimation error, but recovering the plant state remains challenging, as both the controller's state and input remain encrypted.
\end{rem}

\begin{figure*}[t]
\centering
\input{img/diagram}
\caption{Configuration of the proposed encrypted control system with an anomaly detector. 
The system operator generates and distributes the secret key to the plant side and initializes the system by encrypting the controller's initial state. 
The right panel provides a detailed diagram of $\Enc_t$.}
\label{fig:diagram}
\end{figure*}

\begin{rem}\upshape
    In terms of computational effort, the proposed encryption scheme appends one additional column to each ciphertext, but this incurs a negligible increase as the dimension $N$ is typically chosen as a large number (for example, $N=2^{11}$ in Section~\ref{sec:simul}).
    Meanwhile, considering that the masking terms $b_x^\ini$ and $\{b_y(\tau)\}_{\tau=0}^\infty$ of the standard LWE based encryptions in \eqref{eq:standardEncEx} are independent of the corresponding plaintexts, they can be generated offline prior to receiving the plaintexts. 
    This enables one to prepare $V_2T_2b_x^\ini$ and $b_y'(t)$ in Algorithm~\ref{alg:enc} using \eqref{eq:zeroDynState}, thereby reducing the online computational burden.
\end{rem}

\section{Application to Dynamic Controllers Over $\bbR$}\label{sec:application}

We present a method for applying the proposed encryption scheme to dynamic controllers over $\bbR$.
To encrypt a dynamic controller, it is well known that the state matrix of the controller needs to be an integer matrix \cite{CheoHank18}.
Unlike the approaches in \cite{KimjShim23,TeraSada23,LeejLeed24} that re-encrypt the controller output to convert the state matrix into an integer matrix, our method reuses the disclosed residue signal as a fed-back input, thereby reducing communication overhead (see Fig.~\ref{fig:diagram}).

Consider a discrete-time single-input single-output plant written by 
\begin{equation}\label{eq:plant}
    \begin{split}
        x_p(t+1) &= A_px_p(t) + B_pu(t), \quad x_p(0) = x_{p}^\ini, \\
        y(t)  &= C_px_p(t),
    \end{split}
\end{equation}
where $x_p(t)\in\bbR^n$ is the state with the initial value $x_{p}^\ini\in\bbR^n$, $u(t)\in\bbR$ is the input, and $y(t)\in\bbR$ is the output.
Suppose that a controller that stabilizes \eqref{eq:plant} has been designed as
\begin{align}\label{eq:controller}
    x(t+1) &= Ax(t) + By(t), \quad x(0) = x^\ini, \nonumber\\
    u(t) &= Cx(t), \\
    r(t) &= Dx(t)+Ey(t),\nonumber
\end{align}
where $x(t) \in \bbR^n$ is the state with the initial value $x^\ini\in\bbR^n$ and $r(t)\in\bbR$ is the residue signal for anomaly detection.
The matrices in \eqref{eq:plant} and \eqref{eq:controller} consist of real numbers unlike \eqref{eq:sys}.

The objective is to design an encrypted controller that performs the operations of \eqref{eq:controller} using the proposed encryption scheme, which ensures the followings: 
Let us denote the plant input and the residue signal of the closed-loop system \eqref{eq:plant} with \eqref{eq:controller} by $u^\nom(t)$ and $r^\nom(t)$, respectively. 
Then, for given $\epsilon>0$,
\begin{itemize}[leftmargin=*]
    \item the encrypted controller automatically discloses a residue signal $r(t)$ without decryption such that
    \begin{equation}\label{eq:goalr}
        \left\|r(t) - r^\nom(t) \right\|\le \epsilon
    \end{equation}
    for all $t\ge 0$;
    \item the control performance of the encrypted controller is equivalent to that of the controller \eqref{eq:controller} in the sense that 
    \begin{equation}\label{eq:goalu}
        \left\|u(t) - u^\nom(t) \right\|\le \epsilon
    \end{equation}
    for all $t\ge 0$, where $u(t)$ is the control input decrypted from the encrypted controller.
\end{itemize}

\subsection{Conversion to system over $\bbZ_q$}\label{subsec:sysZq}

We begin by converting the controller \eqref{eq:controller} to operate over the plaintext space $\bbZ_q$.
To this end, we adopt the approach of \cite{KimjShim23} and first convert the state matrix of \eqref{eq:controller} into an integer matrix via pole-placement.
For this, we introduce the following assumption.
\begin{asm}\upshape\label{asm:obsv}
    The pair $(A,D)$ is observable.
\end{asm}

Under this assumption, there exist a vector $Q\in\bbR^n$ and an invertible matrix $T\in \bbR^{n\times n}$ such that 
\begin{equation}\label{eq:defF}
    F= T(A-QD)T^{-1}\in \bbZ^{n \times n},
\end{equation}
where the eigenvalues of $F$ can be arbitrarily assigned through the design of $Q$.
Here, we specifically assign all eigenvalues of $F$ at the origin, and choose $T$ such that $F$ is in the observable canonical form, written by
\begin{equation}\label{eq:nilpotent}
    F = \begin{bmatrix}
        0 & \cdots & 0 & 0 \\
        1 & \cdots & 0 & 0 \\
        \vdots & \ddots & \vdots & \vdots \\
        0 & \cdots & 1 & 0
    \end{bmatrix}.
\end{equation}
Then, $F$ is nilpotent of index $n$, i.e., $F^n=\bfzero_{n\times n}$, and this will become relevant in the next subsection when analyzing the performance of the resulting encrypted controller.

With the coordinate transformation $z(t) = Tx(t)$, the controller \eqref{eq:controller} can be rewritten as 
\begin{align}\label{eq:ctrl_intState}
    z(t+1) &= Fz(t) + T(B\!-\!QE)y(t) + TQr(t), ~  z(0) \!=\!  Tx^\ini, \nonumber\\
    u(t) &= CT^{-1}z(t), \\
    r(t) &= DT^{-1}z(t) + Ey(t),\nonumber
\end{align}
where the integer matrix $F$ can be considered as the state matrix, regarding $r(t)$ as a fed-back input with the gain $TQ$.

Next, we convert \eqref{eq:ctrl_intState} to operate over $\bbZ_q$.
To reduce precision loss and preserve the significand of the fractional components, we scale and round the matrices in \eqref{eq:ctrl_intState}, except for $F$ which is already an integer matrix, using a scale factor $1/\sfs_1\in \bbN$.
This yields the following integer-valued matrices:
\begin{subequations}\label{eq:ctrlZ_q}
\begin{align}\label{eq:paramsZ_q}
     G&=\left\lceil \frac{T(B-QE)}{\sfs_1} \right\rfloor, &
    R&=\left\lceil \frac{TQ}{\sfs_1} \right\rfloor, & 
    P&=\left\lceil \frac{CT^{-1}}{\sfs_1} \right\rfloor, \nonumber \\
    H&=\left\lceil \frac{DT^{-1}}{\sfs_1} \right\rfloor, & J&=\llceil \frac{E}{\sfs_1^2}\rrfloor.
\end{align}
Correspondingly, we let the initial value $Tx^\ini$ and the plant output $y(t)$ (for each $t\ge 0$) be quantized as
\begin{equation}\label{eq:signalsZ_q}
    \tilde{x}_{q}^\ini = \left \lceil \frac{Tx^\ini}{\sfs_1\sfs_2} \right\rfloor\modp q, ~~ 
    \tilde{y}_q(t) =  \left \lceil \frac{y(t)}{\sfs_2} \right \rfloor\modp q,
\end{equation}
where $\sfs_2>0$ denotes the step size for quantization.
For further details regarding this quantization procedure and its implications, we refer the reader to \cite{KimjShim23}.

As a result, we obtain a controller that operates over $\bbZ_q$:
\begin{align}\label{eq:stateDynRe}
    \tilde{x}_q(t+1) &= F\tilde{x}_q(t) + G\tilde{y}_q(t)+R\hat{r}_q(t)\modp q, \nonumber\\
    \tilde{u}_q(t) &= P\tilde{x}_q(t)\modp q, \\
    \tilde{r}_q(t) &= H\tilde{x}_q(t) + J\tilde{y}_q(t)\modp q, \nonumber \\
    \tilde{x}_q(0) &= \tilde{x}_{q}^\ini, \nonumber
\end{align}
where $\tilde{x}_q(t)\in\bbZ_q^n$, $\tilde{u}_q(t)\in\bbZ_q$, and $\tilde{r}_q(t)\in\bbZ_q$ are the state, the output, and the residue signal, respectively. 
The fed-back input $\hat{r}_q(t)\in\bbZ_q$ is defined by
\begin{equation}\label{eq:hatrq}
        \hat{r}_q(t)=\sfQ(\tilde{r}_q(t)) :=
        \left\lceil \sfs_1^2\cdot \tilde{r}_q(t) \right\rfloor .
\end{equation}

The rationale behind scaling $\tilde{r}_q(t)$ by $\sfs_1^2$ prior to feedback is that $\tilde{x}_q(t)$ and $\tilde{r}_q(t)$ are of scale $1/\sfs_1\sfs_2$ and $1/\sfs_1^2\sfs_2$, respectively, i.e., they have approximate values of
\begin{equation*}
    \tilde{x}_q(t)\approx \frac{z(t)}{\sfs_1\sfs_2} \modp q, \quad  \tilde{r}_q(t) \approx \frac{r(t)}{\sfs_1^2\sfs_2} \modp q.
\end{equation*}
Since the matrix $R$ is also of scale $1/\sfs_1$ with respect to the matrix $TQ$, we re-scale $\tilde{r}_q(t)$ into a scale of $1/\sfs_2$, so that the scale of $R\hat{r}_q(t)$ matches that of $\tilde{x}_q(t)$.
By a similar reasoning, we let the plant input $u(t)$ and the residue signal $r(t)$ for anomaly detection be obtained from \eqref{eq:stateDynRe} as 
\begin{equation}\label{eq:ctrlZ_qInputResidue}
        u(t) = \sfs_2 \cdot \sfQ(\tilde{u}_q(t)), ~~~~~
        r(t)= \sfs_2\cdot \sfQ(\tilde{r}_q(t)).
    \end{equation}
\end{subequations}

The following lemma states that $u(t)$ and $r(t)$ generated by \eqref{eq:ctrlZ_qInputResidue} can be made arbitrarily close to $u^\nom(t)$ and $r^\nom(t)$, respectively, by selecting sufficiently small scale parameters $\{\sfs_1,\sfs_2\}$ and a sufficiently large modulus $q$.

\begin{lem}
\upshape\label{lem:equivZq} 
    Assume that the closed-loop system \eqref{eq:plant} with \eqref{eq:controller} is stable.
    For a given $\epsilon>0$, there exist continuous functions $\beta(\sfs_1,\sfs_2)$ 
    and $\gamma(\sfs_1,\sfs_2)$ 
    vanishing at the origin such that if
    \begin{subequations}\label{eq:paramsCond}
        \begin{align}
            \beta(\sfs_1,\sfs_2) &\le \epsilon, \label{eq:beta}\\
            q&>\frac{1}{\gamma(\sfs_1,\sfs_2)},\label{eq:gamma}
        \end{align}
    \end{subequations}
    then the controller \eqref{eq:ctrlZ_q} guarantees that \eqref{eq:goalr} and \eqref{eq:goalu} hold for all $t\ge 0$. 
\end{lem}
\begin{proof}
    The result follows directly by applying Proposition~6 and Theorem~1 of \cite{KimjShim23}, originally derived for a general nonlinear plant, and is here specialized to the linear plant \eqref{eq:plant}.
    For detailed derivations, we refer the reader to \cite{KimjShim23}.
\end{proof}

Note that the condition \eqref{eq:beta} can always be met by decreasing the parameters $\{\sfs_1,\sfs_2\}$, which reduces the precision losses caused by the rounding operations in \eqref{eq:paramsZ_q} and \eqref{eq:signalsZ_q}.
However, decreasing $\sfs_1$ and $\sfs_2$ also increases the scale of the signals $\tilde{u}_q(t)$ and $\tilde{r}_q(t)$.
If the modulus $q$ is not sufficiently large, 
the higher bits of $\tilde{u}_q(t)$ and $\tilde{r}_q(t)$ may be truncated by the modulo operations in \eqref{eq:stateDynRe}.
The condition \eqref{eq:gamma} serves to prevent this issue by ensuring that $q$ is sufficiently large to cover the range of these signals. 
To be precise, it guarantees that $\tilde{u}_q(t)=P\tilde{x}_q(t)$ and $\tilde{r}_q(t)=H\tilde{x}_q(t)+J\tilde{y}_q(t)$, i.e., the modulo operation can be omitted.
Therefore, the signals remain bounded as
\begin{equation}\label{eq:uqrqbound}
    \left\|
    \begin{bmatrix}
        \tilde{u}_q(t) \\ 
        \tilde{r}_q(t)
    \end{bmatrix} \right\| \le \frac{1}{2\gamma(\sfs_1,\sfs_2)} < \frac{q}{2}
\end{equation}
for all $t\ge 0$.
While a detailed derivation is omitted due to space limitations, it can be easily derived from \cite{KimjShim23}.

\subsection{Encrypted controller design and performance analysis}\label{subsec:controllerDesign}

We present a method to apply the proposed encryption scheme to the converted controller \eqref{eq:ctrlZ_q} over $\bbZ_q$.
The key idea is to directly reuse the disclosed residue signal for constructing the fed-back input term in the state dynamics, instead of re-encrypting the residue signal at the actuator.

We propose an encrypted controller of the form \eqref{eq:sysEncEx2}, with its state dynamics slightly modified:
\begin{subequations}\label{eq:encController}
\begin{align}\label{eq:encControllerState}
    \bfx(t+1) &= F \cdot \bfx(t) + G \cdot \Enc_t(y_q(t)) + R \cdot \hat{\bfr}(t) \modp q, \nonumber \\
        \bfu(t) &= P\cdot \bfx(t) \modp q, \nonumber \\
        \bfr(t) &= H\cdot \bfx(t) + J \cdot \Enc_t(y_q(t)) \modp q, \\
        \bfx(0) &= \Enc_\ini(x_{q}^\ini),\nonumber
\end{align}
where 
\begin{equation}\label{eq:scaledInRes}
    y_q(t)=\frac{\tilde{y}_q(t)}{\sfL}\modp q,~~~~~x_q^\ini = \frac{\tilde{x}_q^\ini}{\sfL} \modp q.
\end{equation}
Here, the scale factor $1/\sfL\in\bbN$ is introduced to negate the effect of error terms injected during encryption, as discussed in Remark~\ref{rem:L}.
The fed-back input $\hat{\bfr}(t)\in \bbZ_q^{1\times(N+2)}$ is defined by
\begin{equation}\label{eq:encControllerFeedback}
    \hat{\bfr}(t) :=
    \frac{1}{\sfL}\cdot \begin{bmatrix}
        \sfQ(\sfL\cdot \bfr_1(t)), &\!\! \bfzero_{1\times(N+1)}
    \end{bmatrix},
\end{equation}
where $\bfr_1(t)$ denotes the first element of $\bfr(t)$, as in \eqref{eq:bfr}.

The output $\bfu(t)$ is transmitted to the plant, decrypted, and then scaled down to obtain the plant input, as 
\begin{equation}\label{eq:encControllerInput}
    u(t) = \sfs_2 \cdot\sfQ\left( \left\lceil\sfL\cdot\Dec'(\bfu(t))\right\rfloor \right).
\end{equation}
The residue signal $r(t)$ for anomaly detection is obtained from $\bfr(t)$ without decryption, as 
\begin{equation}\label{eq:encControllerResidue}
    r(t) =  \sfs_2 \cdot \sfQ(\sfL\cdot\bfr_1(t)).
\end{equation}
\end{subequations}
A complete configuration of the proposed encrypted control with \eqref{eq:plant} and \eqref{eq:encController} is depicted in Fig.~\ref{fig:diagram}.

The following theorem states that the proposed encrypted controller ensures \eqref{eq:goalr} and \eqref{eq:goalu} for all $t\ge0$ with appropriate choice of the parameters $\{\sfs_1,\sfs_2,\sfL\}$ and the modulus $q$.

\begin{thm}\label{thm:performance}\upshape
    Assume that the closed-loop system \eqref{eq:plant} with \eqref{eq:controller} is stable, and define $\sfM:=\left\| P \right\| \left(1+n\cdot \left\| G\right\| \right)\delta$.
    For given $\epsilon>0$, if the parameters $1/\sfs_1\in\bbN$, $\sfs_2>0$, and $1/\sfL\in\bbN$, and the modulus $q>0$ satisfy \eqref{eq:paramsCond},
    \begin{subequations}\label{eq:thmFinalC}
    \begin{align}
        \sfL\sfM &< \frac{1}{2},\label{eq:thmFinalCondL} \\
        q &> \frac{1}{\sfL\cdot \gamma(\sfs_1,\sfs_2)} + 2\sfM, \label{eq:thmFinalCond} 
    \end{align}
    \end{subequations}
    then the encrypted controller \eqref{eq:encController} guarantees that \eqref{eq:goalr} and \eqref{eq:goalu} hold for all $t\ge 0$.
\end{thm}

Before proving the theorem, we provide the rationale behind the construction of the fed-back input $\hat{\bfr}(t)$ in \eqref{eq:encControllerState}, and show that the error terms injected during encryption does not affect the performance of the encrypted controller.

First, note that $\hat{\bfr}(t)$ can be interpreted as a ciphertext with the plaintext $\sfQ(\sfL\cdot \bfr_1(t))/\sfL$, whose random matrix and masking terms are both zero.
This implies that the masking terms of $\bfx(t)$ and $\bfr(t)$ of \eqref{eq:encControllerState} still evolve according to the dynamics \eqref{eq:bDyn}.
Therefore, even though the proposed encryption scheme does not explicitly account for the additional term $R\cdot\hat{\bfr}(t)$ in \eqref{eq:encControllerState}, the plaintext of $\bfr(t)$ is nonetheless correctly disclosed as $\bfr_1(t)$ by Theorem~\ref{thm:encryption}.
Consequently, the dynamics governing the plaintexts in \eqref{eq:encControllerState} admits the following representation, similar to \eqref{eq:stateDynRe}:
\begin{align}\label{eq:encMsg}
    x_q(t+1) &= Fx_q(t) + Gy_q(t) + R\cdot \frac{\sfQ(\sfL\cdot r_q(t))}{\sfL} \modp q, \nonumber\\
    u_q(t) &= Px_q(t) \modp q, \\
    r_q(t) &= Hx_q(t) + J y_q(t) \modp q,\nonumber \\
    x_q(0) &= x_q^\ini , \nonumber
\end{align}
where $x_q(t)\in\bbZ_q^n$, $u_q(t)\in\bbZ_q$, and $r_q(t)\in\bbZ_q$ are the plaintexts of the ciphertexts $\bfx(t)$, $\bfu(t)$, and $\bfr(t)$ in \eqref{eq:encControllerState}, respectively.
In particular, $\bfr_1(t)$ in \eqref{eq:encControllerFeedback} coincides with $r_q(t)$ in \eqref{eq:encMsg}.

Next, we examine the growth of the error terms injected during encryption.
Let $e_u(t)\in\bbZ$ denote the error term of the ciphertext $\bfu(t)$, which satisfies
\begin{equation*}
    \Dec'(\bfu(t))=u_q(t) + e_u(t) \modp q.    
\end{equation*}
As can be seen from \eqref{eq:encControllerInput}, this error term is recovered along with the plaintext during decryption and may propagate through the plant input, affecting the overall behavior of the closed-loop system.

This is precisely where the nilpotency of $F$ becomes relevant.
Specifically, it guarantees that $e_u(t)$, which is generated through \eqref{eq:encControllerState}, remains bounded and does not overflow the range of $\bbZ_q$. 
Given that the error term is bounded, its effect can be entirely eliminated by the rounding operation in \eqref{eq:encControllerInput}, provided that the scaling factor $\sfL$ is chosen sufficiently small.

To analyze the boundedness of $e_u(t)$, observe that it obeys the following dynamics over $\bbZ$:
\begin{align}\label{eq:errorDyn}
    e_x(t+1)&=Fe_x(t) +Ge_y(t), ~~~ e_x(0)=e_x^\ini, \\
    e_u(t) &= Pe_x(t), \nonumber
\end{align}
where $e_x^\ini$ and $e_y(t)$ are defined as in \eqref{eq:standardEncExB}. 
This follows directly from the linearity of \eqref{eq:encControllerState} and the fact that $\hat{\bfr}(t)$ can be regarded as a ciphertext with a zero error term.
Since $F$ is nilpotent of index $n$ and is given in the observable canonical form, we have
\begin{align}\label{eq:sfM}
    &\|e_u(t)\| = \left\| P\left(F^t e_x^\ini + \sum_{k=0}^{t-1}F^kGe_y(t-1-k) \right) \right\| \nonumber \\
    &\le \left\| P \right\|\left(\left\|F^t \right\| + \sum_{k=0}^{t-1}\left\|F^k \right\|\cdot \left\| G \right\| \right) \delta \nonumber \\
    &\le \left\| P \right\| \left(1+n\cdot \left\| G\right\| \right)\delta=\sfM.
\end{align}
Under \eqref{eq:sfM}, it is expected that the effect of this bounded error term can be removed by the rounding operation in \eqref{eq:encControllerInput}.
We now proceed to the proof of Theorem~\ref{thm:performance}.

\begin{proof}[Proof of Theorem~\ref{thm:performance}]
    Consider the closed-loop of \eqref{eq:plant} with the controller \eqref{eq:ctrlZ_q} over $\bbZ_q$, and that of \eqref{eq:plant} with the encrypted controller \eqref{eq:encController}.
    We show that both controllers generate identical control inputs $u(t)$ and residue signals $r(t)$ for all $t\ge 0$ in their respective closed-loop systems.

    Without loss of generality, assume that the controllers \eqref{eq:ctrlZ_q} and \eqref{eq:encController} receive the same input $y_q(0)$ at $t=0$.
    Then, it follows from \eqref{eq:ctrlZ_q}, \eqref{eq:scaledInRes}, and \eqref{eq:encMsg} that 
    \begin{align*}
        u_q(0)=\frac{\tilde{u}_q(0)}{\sfL} \modp q, ~~~~
        r_q(0)=\frac{\tilde{r}_q(0)}{\sfL} \modp q.
    \end{align*}
    Under the condition \eqref{eq:thmFinalCond}, it is ensured by \eqref{eq:uqrqbound} that
    \begin{equation*}
        \left\| \begin{bmatrix}
            \tilde{u}_q(0)/\sfL \\
            \tilde{r}_q(0)/\sfL
        \end{bmatrix}\right\| < \frac{q}{2},
    \end{equation*}
    which implies that the modulo operation can be omitted, i.e.,
    \begin{equation}\label{eq:auxProp}
        \bfr_1(0)=r_q(0)=\frac{\tilde{r}_q(0)}{\sfL}, ~~~u_q(0)=\frac{\tilde{u}_q(0)}{\sfL}.
    \end{equation}
    Consequently, the residue signals $r(0)$ obtained from \eqref{eq:ctrlZ_qInputResidue} and  \eqref{eq:encControllerResidue} are identical.
    Similarly, the control inputs $u(0)$ computed from \eqref{eq:ctrlZ_qInputResidue} and \eqref{eq:encControllerInput} are also identical because
    \begin{align*}
        \left\lceil \sfL\cdot \Dec'(\bfu(0))\right\rfloor &= \left\lceil \sfL \cdot \left(u_q(0) + e_u(0)  \modp q\right)\right \rfloor \\
        &= \left\lceil \sfL \cdot \left(\frac{\tilde{u}_q(0)}{\sfL} + e_u(0) \right)\right \rfloor = \tilde{u}_q(0),
    \end{align*}
    where the second equality holds because $\|u_q(0) + e_u(0) \| < q/2$ by \eqref{eq:uqrqbound}, \eqref{eq:thmFinalCond}, and \eqref{eq:sfM}, and the last equality holds because $\|\sfL\cdot e_u(0)\|\le \sfL\sfM<1/2$.
    Lastly, it follows from \eqref{eq:scaledInRes} and the identity $\sfQ(\sfL\cdot r_q(0))/\sfL = \hat{r}(0)/\sfL$ that $x_q(1)=\tilde{x}_q(1)/\sfL\modp q$.

    Now suppose that for some $k\ge 1$, the control inputs $u(t)$ and residue signals $r(t)$ obtained from \eqref{eq:ctrlZ_qInputResidue}, \eqref{eq:encControllerInput}, and \eqref{eq:encControllerResidue} are identical, and $x_q(t+1)=\tilde{x}_q(t+1)/\sfL\modp q$ hold for $t=0,1,\ldots,k-1$. 
    Then, the two controllers receive the same input $y_q(k)$ at $t=k$ in their respective closed-loop systems. 
    By applying the same reasoning as in the case $t=0$, it is obtained that 
    \begin{align*}
        \bfr_1(k)=\frac{\tilde{r}_q(k)}{\sfL}, ~~~~
        \llceil \sfL\cdot \Dec'(\bfu(k)) \rrfloor = \tilde{u}_q(k),
    \end{align*}
    so that $u(k)$ and $r(k)$ obtained by \eqref{eq:ctrlZ_qInputResidue}, \eqref{eq:encControllerInput}, and \eqref{eq:encControllerResidue} are again identical, and $x_q(k+1)=\tilde{x}_q(k+1)/\sfL\modp q$.

    By induction, we conclude that the encrypted controller \eqref{eq:encController} generates the same control inputs and residue signals as the controller \eqref{eq:ctrlZ_q} in the closed-loop with \eqref{eq:plant}. 
    Therefore, Lemma~\ref{lem:equivZq} ensures that \eqref{eq:goalr} and \eqref{eq:goalu} hold for all $t\ge 0$, and this concludes the proof. 
\end{proof}

Note that \eqref{eq:thmFinalC} can always be met by choosing sufficiently small $\sfs_1$, $\sfs_2$, and $\sfL$, and a sufficiently large $q$.
As a practical guideline, we suggest choosing $\sfs_1$, $\sfs_2$, and $\sfL$ as large as possible first while satisfying \eqref{eq:beta} and \eqref{eq:thmFinalCondL} because choosing them excessively small can significantly increase the lower bound \eqref{eq:thmFinalCond} of the modulus $q$, which may lead to higher computational cost.
In particular, we recommend prioritizing a large $\sfs_2$, since decreasing $\sfs_2$ may demand higher sensor resolution in practice. 
Then, select $q$ to satisfy \eqref{eq:thmFinalCond}.

\begin{rem}\upshape
    The proposed encrypted controller \eqref{eq:encController} is capable of operating for an infinite time horizon without re-encryption, unlike the previous result \cite{KimjShim23}.
    Re-encryption requires an additional communication link between the actuator and the encrypted controller, since the controller output is decrypted, re-scaled, encrypted, and then transmitted back to the controller.
    Instead, we utilized the disclosed residue signal to convert the state matrix of the given controller into an integer matrix,
    and re-scaled it directly, as in \eqref{eq:encControllerResidue}.
    Therefore, it can be implemented without an additional communication link with the actuator,
    reducing both the computation time and communication burden.
\end{rem}

\begin{figure}[t]
\centering
\begin{tikzpicture}[x=1cm,y=1cm,>=latex,thick]

\def\yG{-0.8}      
\def\mW{1.6}       
\def\mH{1.6}       
\def\gap{1}      
\def\xA{0}         
\def\xB{4.0}       

\node[draw,minimum width=\mW cm,minimum height=\mH cm,inner sep=0pt] (m1) at (\xA,0) {$m_1$};
\node[draw,minimum width=\mW cm,minimum height=\mH cm,inner sep=0pt] (m2) at (\xB,0) {$m_2$};

\draw (-1,\yG) -- (5,\yG);

\foreach \x in {-1,-0.6,...,5} {
  \draw (\x,\yG) -- ++(-0.35,-0.35);
}

\draw[->,line width=1.2pt] (-1.6,0) -- ($(m1.west)+(-0.15,0)$);
\node at (-1.3,0.45) {$u(t)$};

\coordinate (A) at ($(m1.east)+(0.0,0)$);
\coordinate (B) at ($(m2.west)+(0.0,0)$);

\draw (A) -- ++(0.6,0) coordinate (A2);
\draw (B) -- ++(-0.6,0) coordinate (B2);

\draw[decorate,decoration={zigzag,segment length=8pt,amplitude=5pt}]
  (A2) -- (B2);

\node[above, yshift=6pt] at ($(A2)!0.5!(B2)$) {$k$};

\end{tikzpicture}
\caption{Configuration of the two-mass-spring system.} 
\label{fig:cart}   
\end{figure}

\section{Numerical Simulations}\label{sec:simul}

This section provides simulation results of the proposed method applied to a two-mass-spring system \cite{WiebBern92}, depicted in Fig.~\ref{fig:cart}.
The model of the form \eqref{eq:plant} is obtained as 
\begin{align*}
    A_p &= 
    \begin{bmatrix}
     0.9950, & 0.0998, & 0.0050, & 0.0002 \\
   -0.0997, & 0.9950, & 0.0997, & 0.0050 \\
    0.0050, & 0.0002, & 0.9950, & 0.0998 \\
    0.0997, & 0.0050, & -0.0997, & 0.9950
    \end{bmatrix}, \\ 
    B_p &= 
    \begin{bmatrix}
    0.0050 \\
    0.0998 \\
         0 \\
    0.0002 \\
    \end{bmatrix}, \quad C_p=\begin{bmatrix}
        0, & 0, & 1, & 0
    \end{bmatrix},
\end{align*}
by discretizing the system with the sampling period of $\SI{0.1}{\s}$, where the parameters are set as $m_1=m_2=\SI{1}{\kg}$ and $k=\SI{2}{\newton/\meter}$.
The state $x_p(t)=:[x_{p,1}(t);x_{p,2}(t);x_{p,3}(t);x_{p,4}(t)]$ consists of the positions and velocities of the masses, with $x_{p,1}(t)$ and $x_{p,3}(t)$ as the positions, and $x_{p,2}(t)$ and $x_{p,4}(t)$ as the velocities of the left and right masses, respectively.

Let the controller \eqref{eq:controller} be designed as an observer based controller with
\begin{align*}
    A&=A_p+B_pK-LC_p, & B&=L, & C&=K, \\
    D&=-C_p, & E&=1,
\end{align*}
where the state feedback gain $K\in\bbR^{1\times 4}$ and the observer gain $L\in\bbR^{4}$ are given by 
\begin{align*}
    K&=\begin{bmatrix}
         -4.7413, & -3.9785, & 1.2030, & -2.9269
    \end{bmatrix}, \\
    L&=\begin{bmatrix}
        1.0387, & -0.4317, & 1.0914, & 1.6131
    \end{bmatrix}^\top,
\end{align*}
satisfying Assumption~\ref{asm:obsv}. 
The matrices $Q\in\bbR^4$ and $T\in\bbR^{4\times 4}$ that yield \eqref{eq:nilpotent} are found as below, following the method of \cite[Lemma~1]{KimjShim23}:
\begin{align*}
    Q &=
    \begin{bmatrix}
        -88.3967 &
        -43.7434 &
        -2.4071 &
        -31.8077
    \end{bmatrix}^\top, \\
    T &=
    \begin{bmatrix}
        -79.1486, & 1.5913, & 51.2682, & 78.2571 \\
  984.3034, & 627.5648, & 364.2297, & 174.7009 \\
         0, & 0, & 0, & 1 \\
    4.9769, & -2.0480, & 3.7919, & 17.0544 
    \end{bmatrix}.
\end{align*}

For the simulation, we fixed the encryption parameters as $(N,q,\sigma)=(2^{11},72057594037927931,3.2)$, where $q\approx2^{56}$, ensuring $128$-bit security \cite{AlbrChas21}. 
The bound for the error distribution is set as $\delta=6\sigma$, and the scaling parameters are chosen as $\sfs_1=\sfs_2=\sfL=10^{-4}$.
The initial values for the plant \eqref{eq:plant} and the controller \eqref{eq:controller} are given by $x_{p}^\ini=[1;1;1;1]$ and $x^\ini=[0;0;0;0]$.

We consider an attack scenario, in which an adversary injects an additive attack signal $a(t)\in\bbR$ into the sensor output $y(t)$.
This type of attack has also been studied in \cite{AlexBurb22}, and we adopt this setting to facilitate a clear comparison of the unencrypted controller \eqref{eq:controller} and the encrypted controller \eqref{eq:encController}.

The attack is initiated at $t=50$, as shown in Fig.~\ref{fig:attack}.
To detect the attack, we implement a cumulative sum (CUSUM) based anomaly detector \cite{SandGupt22}.
Given the residue signal $r(t)$ at time step $t$, the CUSUM statistic $S(t)\in\bbR$ is updated as
\begin{align}\label{eq:CUSUM}
    S(t+1) = \max\{S(t)+|r(t)|^2-\alpha,0\},
\end{align}
where $S(0)=0$, and $\alpha>0$ is a tunable forgetting factor that determines the sensitivity to past residue signals.
An alarm is triggered whenever 
\begin{equation*}
    S(t)>\eta,
\end{equation*}
for a predefined threshold $\eta>0$.
The forgetting factor and the threshold are set as $\alpha=0.2$ and $\eta =0.1$, respectively.

\begin{figure}[t]
    \begin{centering}
        \input{img/attack}
        \caption{Injected attack signal $a(t)$ beginning at $t = 50$ (\SI{5}{\sec}).} 
        \label{fig:attack}
    \end{centering}
\end{figure}


\begin{figure}[t]
    \begin{centering}
        \input{img/residue}
        \caption{Comparison of the residue signal $r(t)$ and CUSUM statistic $S(t)$ obtained from the unencrypted controller \eqref{eq:controller} (blue solid line) and the encrypted controller \eqref{eq:encController} (yellow solid line). The detection threshold $\eta$ is shown as a red dotted line.} 
        \label{fig:CUSUM}
    \end{centering}
\end{figure}

Fig.~\ref{fig:CUSUM} compares the residue signal $r(t)$ and the CUSUM statistic $S(t)$ computed by the unencrypted controller \eqref{eq:controller} and the encrypted controller \eqref{eq:encController}.
The two controllers exhibit comparable performance and both successfully detect the injected attack, which validates the effectiveness of the proposed method. 
The temporary false alarm observed at the beginning is due to transient errors, and may be alleviated through a further tuning of $\alpha$ and $\eta$.

\section{Conclusion}\label{sec:conclusion}
In this paper, we have proposed a homomorphic encryption scheme for dynamic controllers that automatically discloses the residue signal for anomaly detection. 
This enables a network-side detector to directly detect anomalies, without requiring access to the secret key.
The proposed scheme leverages the controller’s zero-dynamics to enforce the masking term of the encrypted residue to remain identically zero, leading to the disclosure of its plaintext. 
It has been shown that the proposed scheme is secure in the sense that it does not compromise the security of the standard LWE based scheme beyond disclosing the residue signal.
Furthermore, we have demonstrated a method to implement dynamic feedback controllers over $\bbR$ using the proposed encryption scheme. 
Our design utilizes the disclosed residue signal as a fed-back input to convert the state matrix of a given controller into an integer matrix, thereby eliminating the need for re-encryption.
 
We reported initial results on extending the proposed framework to multi-input multi-output systems \cite{JangLees25}, and future work will focus on  encrypting the controller parameters as well to further enhance security.

\bibliographystyle{IEEEtran}
\bibliography{ref} 

\vskip -2.5\baselineskip plus -1fil

\begin{IEEEbiography}[{\includegraphics[width=1in,height=1.25in,clip,keepaspectratio]{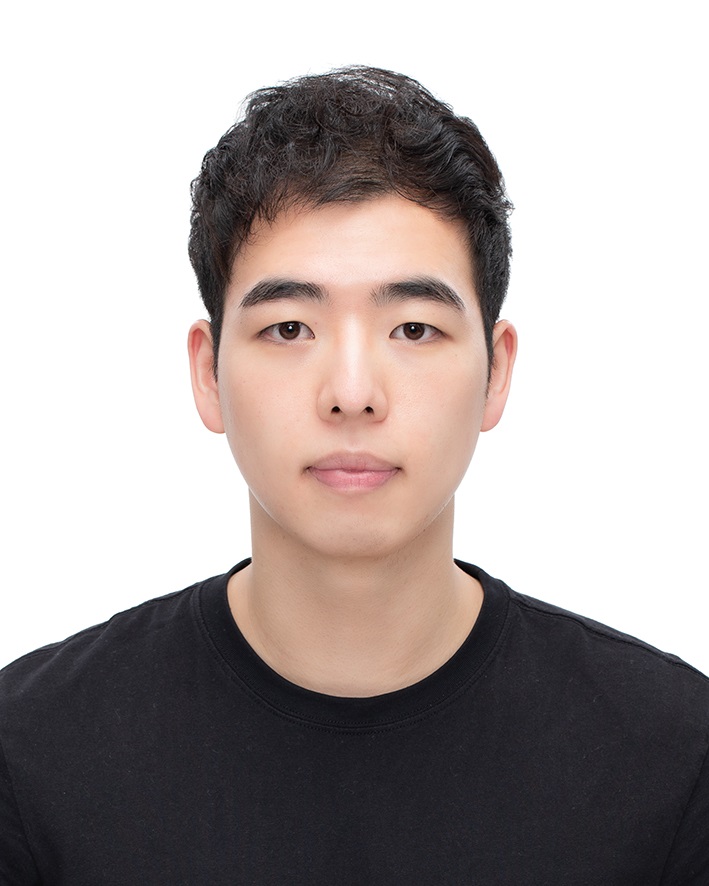}}]%
{Yeongjun Jang}
received the B.S. degree in electrical and computer engineering in 2022, from Seoul National University, South Korea.
He is currently a combined M.S./Ph.D. student in electrical and computer engineering at Seoul National University, South Korea. 
His research interests include data-driven control and encrypted control systems.
\end{IEEEbiography}
\vskip -2.5\baselineskip plus -1fil

\begin{IEEEbiography}[{\includegraphics[width=1in,height=1.25in,clip,keepaspectratio]{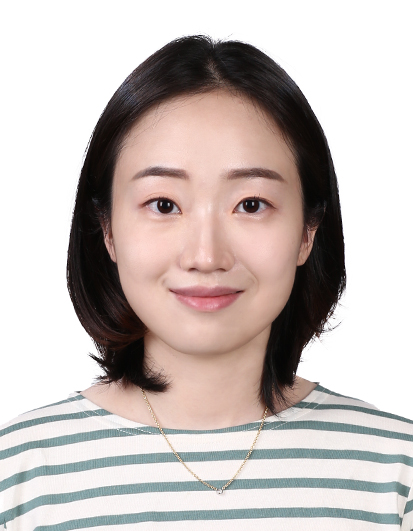}}]%
{Joowon Lee}
received the B.S. and combined M.S./Ph.D. degrees in electrical and computer engineering from Seoul National University, Seoul, South Korea, in 2019 and 2026, respectively.
She is currently a Postdoctoral Researcher with Department of Decision and Control Systems, KTH Royal Institute of Technology, Stockholm, Sweden.
Her research interests include data-driven control and encrypted control.
\end{IEEEbiography}
\vskip -2.5\baselineskip plus -1fil

\begin{IEEEbiography}[{\includegraphics[width=1in,height=1.25in,clip,keepaspectratio]{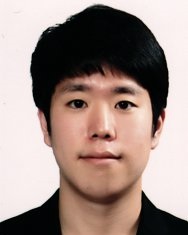}}]%
{Junsoo Kim}
received the B.S. degrees in electrical engineering and mathematical sciences in
 2014, and the M.S. and Ph.D. degrees in electrical engineering in 2020, from Seoul National University, South Korea, respectively. 
 He held the Postdoc position at KTH Royal Institute of Technology, Sweden, till 2022. 
  He is currently an
 Assistant Professor at the Department of Electrical and Information Engineering, Seoul National University of Science and Technology, South Korea. His research interests include security
 problems in networked control systems and encrypted control systems.
\end{IEEEbiography}
\vskip -2.5\baselineskip plus -1fil
\begin{IEEEbiography}[{\includegraphics[width=1in,height=1.25in,clip,keepaspectratio]{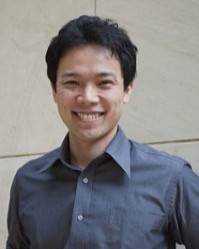}}]%
{Takashi Tanaka}
received the B.S. degree from the University of Tokyo, Tokyo, Japan, in 2006, and the M.S.
and Ph.D. degrees in aerospace engineering (automatic control) from the University of Illinois at Urbana Champaign, Champaign, IL, USA, in 2009 and 2012, respectively. 
He was a Postdoctoral Associate with the Laboratory for Information and Decision Systems at the Massachusetts Institute of Technology, Cambridge, MA, USA, from 2012 to 2015, and a postdoctoral researcher at KTH Royal Institute of Technology, Stockholm, Sweden, from 2015 to 2017. 
He was an Assistant Professor in the Department of Aerospace Engineering and Engineering Mechanics at the University of Texas at Austin between 2017 and 2024, and is an Associate Professor at the School of Aeronautics and Astronautics and the Elmore Family School of Electrical and Computer Engineering at Purdue University since 2025.
His research interests include control theory and its applications, most recently the information-theoretic perspectives of optimal control problems. 
He was the recipient of the DARPA
Young Faculty Award, the AFOSR Young Investigator Program Award, and the NSF Career Award.
\end{IEEEbiography}
\vskip -2.5\baselineskip plus -1fil
\begin{IEEEbiography}[{\includegraphics[width=1in,height=1.25in,clip,keepaspectratio]{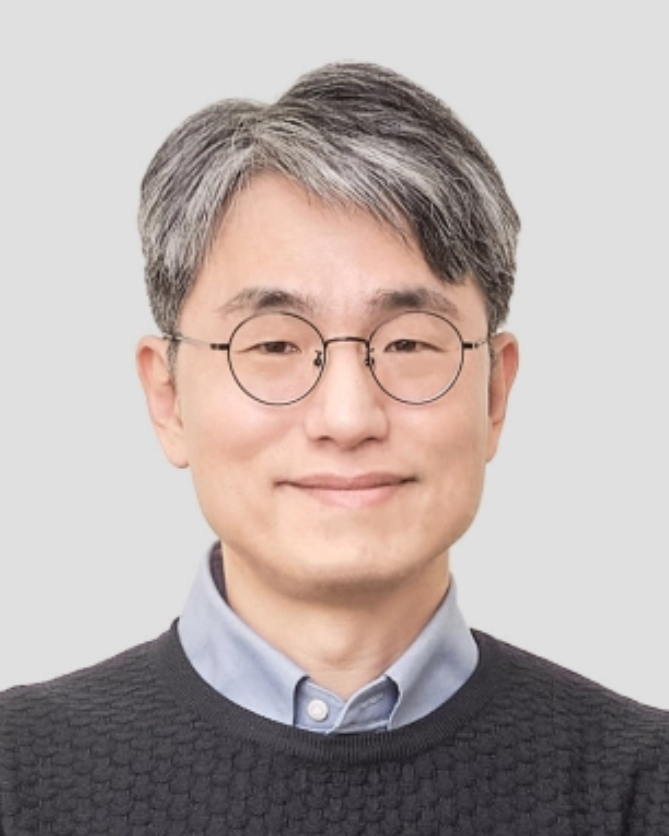}}]%
{Hyungbo Shim} received the B.S., M.S., and Ph.D. degrees from Seoul National University, South Korea. 
He held a postdoctoral research position with the University of California at Santa Barbara, until 2001. 
He joined Hanyang University, Seoul, in 2002. 
Since 2003, he has been with Seoul National University. 
His research interests include stability analysis of nonlinear systems, observer design, disturbance observer, secure control systems, and synchronization in multi-agent systems.
He is a member of Korean Academy of Science and Technology. 
He has served as an Associate Editor for Automatica, IEEE Transactions on Automatic Control, International Journal of
Robust and Nonlinear Control, and European Journal of Control.
\end{IEEEbiography}

\end{document}

%% file: img/diagram.tex
\begin{tikzpicture}[font=\small]

\node[block] (q1) {Scaling \\ {\scriptsize\eqref{eq:encControllerInput}}};
\node[block, right=1.5cm of q1] (plant) {Plant};
\node[block, right=1.5cm of plant] (q2){Quantize\\
{\scriptsize \eqref{eq:signalsZ_q}, \eqref{eq:scaledInRes}} };

\node[block, below=0.8cm of q1] (dec) {$\Dec'$};
\node[block, below=0.8cm of q2] (enc) {$\Enc_t$};

\node[block, below=2.4cm of plant] (oper) {System \\ Operator};

\draw[line, dashed] (oper.north) |- node[midway, above]{$\sk$} (dec.east);

\draw[line, dashed] (oper.north) |- (enc.west);

\draw[line, dashed] (oper.east) -|  node[midway, above, xshift = -0.4cm]{$b_x^\ini$} ($(enc.south)+(-0.3cm,0)$);

\node[draw, 
      minimum width=5.5cm, minimum height=4cm,
      right=2.2cm of enc, align=left] (encZoom) {};

\draw[plain, loosely dashed, thick] (enc.north east) -- (encZoom.north west);
\draw[plain, loosely dashed,thick] (enc.south east) -- (encZoom.south west);


\draw[plain] ($(encZoom.north)+(-16mm,5mm)$) -- node[midway, right]{$y_q(t)$} ($(encZoom.north)+(-16mm,0)$);

\coordinate (BW) at (encZoom.north west);

\node[block, anchor=center] (zEnc)
      at ($(encZoom.north)+(-16mm,-7mm)$) {$\Enc$};

\node[block, anchor=center] (zZD)
      at ($(encZoom.north)+(12mm,-7mm)$) {Zero-Dynamics\\ \eqref{eq:zeroDynState}
      };

\node[block, anchor=north] (yEquiv)
      at ($(encZoom.north)+(12mm,-2cm)$) {\eqref{eq:lem1_claimInput}
      };

\node[block, anchor=south] (zCrypt)
      at ($(encZoom.south)+(0,2mm)$) {\eqref{eq:enc_modified}};

\draw[line, dashed] ($(zZD.north)+(0,7.5mm)$) -- node[midway, right, yshift=1.5mm]{$b_x^\ini$} (zZD.north);

\draw[line, dashed] ($(zEnc.west)+(-0.7cm,0)$) -- node[midway,above, xshift=0.12cm]{$\sk$} (zEnc.west);

\draw[line] ($(encZoom.north)+(-16mm,0)$) -- (zEnc.north);
\draw[line] ($(zEnc.south)+(-4mm,0)$) |-node[midway, below, xshift=2mm]{$\Enc(y_q(t))$}  (zCrypt.west);           
\draw[line] (zEnc.south) |- node[midway, right,yshift=0.3cm]{$b_y(t)$}  (yEquiv.west);           
\draw[line] (yEquiv.south) |- node[midway, right,yshift=0.4cm]{$b_y'(t)$}  (zCrypt.east); 

\draw[line] (zZD.south) --node[midway, right]{$b_z'(t)$} (yEquiv.north);        

\draw[line] ($(zCrypt.south)$) -- node[midway, right, yshift=-1mm]{$\Enc_t(y_q(t))$} ($(encZoom.south)+(0,-5mm)$);

\node[block, below=4.4cm of plant, minimum width=5.5em] (ctrl) {Encrypted\\Controller};

\node[block, below=5.6cm of q1,xshift=1cm] (q3) {Scaling\\
{\scriptsize\eqref{eq:encControllerResidue}}};
\node[block, right=2.5cm of q3] (fb) {Feedback\\
{\scriptsize\eqref{eq:encControllerFeedback}}
};
\node[block, left=1cm of q3] (det) {Anomaly\\Detector}; 

\draw[line] (q1) -- node[midway, above] {$u(t)$} (plant);
\draw[line] (plant) -- node[midway, above] {$y(t)$} (q2);

\draw[line, dashed] (oper.south) -- node[midway, right,yshift=0.1cm]{$\Enc_\ini(x_q^\ini)$} (ctrl.north);

\draw[line] (dec) -- node[midway, right]{$\Dec'(\bfu(t))$} (q1);

\draw[line] (q2) -- node[midway, right, yshift=0.5mm] {$y_q(t)$} (enc);
\draw[line] ($(enc.south)+(0.3cm,0)$) |- node[midway, above, xshift=-9mm]{$\Enc_t(y_q(t))$} ($(ctrl.east)+(0,1.5mm)$);

\draw[line] (fb.north) |- node[midway, right,yshift=-12pt] {$\hat{\bfr}(t)$}($(ctrl.east)+(0,-1.5mm)$);

\draw[line] (ctrl.west) -| node[midway, right,yshift=8pt, xshift = 18pt] {$\bfu(t)$} (dec.south);

\draw[line]  (ctrl.south) |- node[midway, right,yshift=12pt] {$\bfr_1(t)$} (q3);                  
\draw[line]  (ctrl.south) |- (fb);                  
\draw[line]  (q3) -- node[midway, above] {$r(t)$} (det);                

\begin{pgfonlayer}{background}

\node[draw, dotted, thick, rounded corners,
      fit=(q1)(plant)(q2)(dec)(enc), inner sep=8pt] (boxTop) {};

\node[inner sep=0pt, fit=(ctrl)(det)(q3)(fb)] (boxRestFit) {};

\coordinate (cyberSW) at ($(boxRestFit.south west)+(-8pt,-8pt)$);
\coordinate (cyberNE) at ($(boxRestFit.north east)+(12mm,8pt)$);

\node[draw, dotted, thick, rounded corners,
      fit=(cyberSW)(cyberNE), inner sep=0pt] (boxCyber) {};

\node[anchor=south west]
  at ($(boxTop.north west)+(0pt,1pt)$)
  {\bf Plant side};

\node[anchor=south west]
  at ($(boxCyber.north west)+(0pt,1pt)$)
  {\bf Network};

\end{pgfonlayer}

\node[anchor=south east, inner sep=5pt] (legend)
at ($(boxCyber.south east)+(7.5cm,0.1mm)$) {%
\begin{tabular}{@{}l l l l@{}}
\tikz{\draw[line] (0,0)--(10mm,0);} & Online &\qquad
\tikz{\draw[line, dashed] (0,0)--(10mm,0);} & Setup (once)
\end{tabular}
};
\end{tikzpicture}

%% file: img/attack.tex
%
%
\definecolor{mycolor1}{rgb}{0.00000,0.44700,0.84100}%
\begin{tikzpicture}

\begin{axis}[%
width=0.4\textwidth,
height=0.15\textwidth,
at={(0in,0in)},
scale only axis,
xmin=0,
xmax=100,
xlabel style={font=\color{white!15!black}},
xlabel={Time (sec)},
xtick={0,50,100},
xticklabels={0,5,10},
ymin=-0.5,
ymax=0.5,
ylabel style={font=\color{white!15!black},at={(-0.08,0.5)}},
ylabel={$a(t)$},
ytick={-0.5,0,0.5},
axis background/.style={fill=white},
xmajorgrids,
ymajorgrids,
legend style={legend cell align=left, legend columns=2, at={(1,1)},anchor=north east, draw=white!15!black, font=\tiny},
]
\addplot[ycomb, color=mycolor1, line width=1.2pt, mark size=0.7pt, mark=*, mark options={solid, fill=mycolor1, mycolor1}, forget plot] table[row sep=crcr] {%
1	0\\
2	0\\
3	0\\
4	0\\
5	0\\
6	0\\
7	0\\
8	0\\
9	0\\
10	0\\
11	0\\
12	0\\
13	0\\
14	0\\
15	0\\
16	0\\
17	0\\
18	0\\
19	0\\
20	0\\
21	0\\
22	0\\
23	0\\
24	0\\
25	0\\
26	0\\
27	0\\
28	0\\
29	0\\
30	0\\
31	0\\
32	0\\
33	0\\
34	0\\
35	0\\
36	0\\
37	0\\
38	0\\
39	0\\
40	0\\
41	0\\
42	0\\
43	0\\
44	0\\
45	0\\
46	0\\
47	0\\
48	0\\
49	0\\
50	0\\
51	0.314723686393179\\
52	0.405791937075619\\
53	-0.373013183706494\\
54	0.413375856139019\\
55	0.13235924622541\\
56	-0.40245959500059\\
57	-0.221501781132952\\
58	0.0468815192049838\\
59	0.457506835434298\\
60	0.464888535199277\\
61	-0.342386918322452\\
62	0.470592781760616\\
63	0.457166948242946\\
64	-0.0146243512771588\\
65	0.3002804688888\\
66	-0.358113661372785\\
67	-0.078238717373725\\
68	0.415735525189067\\
69	0.292207329559554\\
70	0.459492426392903\\
71	0.155740699156587\\
72	-0.46428832142581\\
73	0.349129305868777\\
74	0.433993247757551\\
75	0.178735154857773\\
76	0.257740130578333\\
77	0.243132468124916\\
78	-0.107772980465832\\
79	0.155477890177557\\
80	-0.328813312188438\\
81	0.206046088019609\\
82	-0.468167153622579\\
83	-0.22307701503911\\
84	-0.453828609368846\\
85	-0.402868218764152\\
86	0.323457828327293\\
87	0.194828622975817\\
88	-0.182900519939139\\
89	0.450222048838355\\
90	-0.465553919497091\\
91	-0.0612556403436018\\
92	-0.118441542906992\\
93	0.265516788149002\\
94	0.295199901137063\\
95	-0.313127395445621\\
96	-0.0102356042117689\\
97	-0.0544137992891005\\
98	0.146313010111265\\
99	0.209364830858073\\
100	0.254686681982361\\
};
\end{axis}
\end{tikzpicture}%

%% file: img/residue.tex
%
%
\definecolor{mycolor1}{rgb}{0.00000,0.44700,0.74100}%
\definecolor{mycolor2}{rgb}{0.92900,0.69400,0.12500}%
\begin{tikzpicture}

\begin{axis}[%
width=0.4\textwidth,
height=0.15\textwidth,
at={(0in,1.4in)},
scale only axis,
xmin=0,
xmax=100,
xtick={0,50,100},
xticklabels={0,5,10},
ymin=-1,
ymax=1,
ylabel style={font=\color{white!15!black},,at={(-0.09,0.5)}},
ylabel={$r(t)$},
ytick={-1,-0.5,0,0.5,1},
axis background/.style={fill=white},
xmajorgrids,
ymajorgrids,
legend style={legend cell align=left, legend columns=2, at={(1,1)},anchor=north east, draw=white!15!black, font=\tiny},
]
\addplot [color=mycolor1, line width=1.2pt, forget plot]
  table[row sep=crcr]{%
1	1\\
2	-0.0182000000000002\\
3	-0.0764130499999951\\
4	-0.0542885302459903\\
5	-0.0271574668076915\\
6	-0.00188969802910632\\
7	0.0205812985624636\\
8	0.0399406060809591\\
9	0.0560323093550035\\
10	0.0688108385464414\\
11	0.0783258279920569\\
12	0.0847088864150349\\
13	0.0881600233662179\\
14	0.0889339317888309\\
15	0.0873264908608338\\
16	0.0836618190263798\\
17	0.0782801539042972\\
18	0.0715267804056232\\
19	0.0637421737514843\\
20	0.0552534716190354\\
21	0.0463673404969418\\
22	0.0373642564090497\\
23	0.0284941801487784\\
24	0.0199735725271575\\
25	0.0119836661440154\\
26	0.00466988693148163\\
27	-0.00185769888453535\\
28	-0.00752304892344391\\
29	-0.0122820634688126\\
30	-0.0161199049341316\\
31	-0.0190478621608787\\
32	-0.0210998983356063\\
33	-0.0223290128792165\\
34	-0.0228035367853486\\
35	-0.0226034681512642\\
36	-0.021816940617484\\
37	-0.020536902641635\\
38	-0.0188580704633314\\
39	-0.0168742027048652\\
40	-0.0146757301745411\\
41	-0.0123477609132327\\
42	-0.00996846810655022\\
43	-0.00760785737076419\\
44	-0.00532690024772015\\
45	-0.00317701259484848\\
46	-0.00119984996291461\\
47	0.00057261299696304\\
48	0.00211775561701938\\
49	0.00342177546513778\\
50	0.00447898331069868\\
51	0.320014654451953\\
52	0.0597335805588625\\
53	-0.835154324026793\\
54	0.830157329970801\\
55	-0.278090861092467\\
56	-0.55112587666914\\
57	0.25176986410844\\
58	0.345563792869873\\
59	0.427566612468118\\
60	-0.0511763243397274\\
61	-0.88710248907482\\
62	0.844886179388657\\
63	-0.022588649650029\\
64	-0.533061422502086\\
65	0.32186815382407\\
66	-0.65229625870063\\
67	0.343355024585354\\
68	0.55836820522909\\
69	-0.151545922125274\\
70	0.1190137979031\\
71	-0.355771132016202\\
72	-0.636255485280833\\
73	0.897030908076004\\
74	0.100998446060848\\
75	-0.31316162472624\\
76	0.0516249507225846\\
77	-0.0272471821932216\\
78	-0.36372716837786\\
79	0.295436524125351\\
80	-0.464756087825751\\
81	0.587573495951386\\
82	-0.664737494256171\\
83	0.290282530410071\\
84	-0.181992790078427\\
85	0.0944905917879714\\
86	0.762115341954077\\
87	-0.198492297440505\\
88	-0.479474589920357\\
89	0.600054284974103\\
90	-0.989305759397876\\
91	0.410119859626372\\
92	-0.0336707865054691\\
93	0.384524329042468\\
94	-0.01184720224732\\
95	-0.670257439524003\\
96	0.327199658175794\\
97	-0.0100581305303614\\
98	0.217767000334314\\
99	0.0585801765752423\\
100	0.0239231005625107\\
};
\addplot [color=mycolor2, line width=1.2pt, forget plot, dashed]
  table[row sep=crcr]{%
1	1\\
2	-0.0182\\
3	-0.0763\\
4	-0.0542\\
5	-0.0271\\
6	-0.0019\\
7	0.0204\\
8	0.0394\\
9	0.0553\\
10	0.0677\\
11	0.077\\
12	0.083\\
13	0.0862\\
14	0.0866\\
15	0.0847\\
16	0.0807\\
17	0.075\\
18	0.0679\\
19	0.0598\\
20	0.051\\
21	0.0418\\
22	0.0324\\
23	0.0232\\
24	0.0144\\
25	0.0062\\
26	-0.0013\\
27	-0.0082\\
28	-0.0139\\
29	-0.019\\
30	-0.0228\\
31	-0.0258\\
32	-0.0279\\
33	-0.029\\
34	-0.0296\\
35	-0.0291\\
36	-0.0282\\
37	-0.0267\\
38	-0.0247\\
39	-0.0225\\
40	-0.02\\
41	-0.0172\\
42	-0.0144\\
43	-0.0117\\
44	-0.009\\
45	-0.0064\\
46	-0.0039\\
47	-0.0017\\
48	0.0003\\
49	0.0021\\
50	0.0035\\
51	0.3196\\
52	0.0595\\
53	-0.8349\\
54	0.8307\\
55	-0.2774\\
56	-0.5501\\
57	0.2529\\
58	0.3468\\
59	0.429\\
60	-0.0497\\
61	-0.8856\\
62	0.8465\\
63	-0.0211\\
64	-0.5317\\
65	0.3232\\
66	-0.6512\\
67	0.3443\\
68	0.5591\\
69	-0.151\\
70	0.1194\\
71	-0.3555\\
72	-0.6361\\
73	0.897\\
74	0.1006\\
75	-0.3136\\
76	0.0511\\
77	-0.028\\
78	-0.3644\\
79	0.2945\\
80	-0.4658\\
81	0.5866\\
82	-0.6658\\
83	0.2893\\
84	-0.183\\
85	0.0935\\
86	0.7613\\
87	-0.1992\\
88	-0.4799\\
89	0.5998\\
90	-0.9894\\
91	0.4102\\
92	-0.0336\\
93	0.384\\
94	-0.0143\\
95	-0.6755\\
96	0.3183\\
97	-0.0232\\
98	0.2005\\
99	0.0372\\
100	-0.0014\\
};
\end{axis}

\begin{axis}[%
width=0.4\textwidth,
height=0.15\textwidth,
at={(0in,0in)},
scale only axis,
xmin=0,
xmax=100,
xlabel style={font=\color{white!15!black}},
xlabel={Time (sec)},
xtick={0,50,100},
xticklabels={0,5,10},
ymin=0,
ymax=2.5,
ylabel style={font=\color{white!15!black},,at={(-0.09,0.5)}},
ylabel={$S(t)$},
ytick={0, 0.5, 1, 1.5, 2, 2.5},
axis background/.style={fill=white},
xmajorgrids,
ymajorgrids,
legend style={legend cell align=left, legend columns=1, at={(0,1)},anchor=north west, draw=white!15!black, font=\footnotesize},
]
\addplot [color=mycolor1, line width=1.2pt]
  table[row sep=crcr]{%
0	0\\
1	0.8\\
2	0.60033124\\
3	0.406170194210302\\
4	0.209117438726572\\
5	0.00985496672998248\\
6	0\\
7	0\\
8	0\\
9	0\\
10	0\\
11	0\\
12	0\\
13	0\\
14	0\\
15	0\\
16	0\\
17	0\\
18	0\\
19	0\\
20	0\\
21	0\\
22	0\\
23	0\\
24	0\\
25	0\\
26	0\\
27	0\\
28	0\\
29	0\\
30	0\\
31	0\\
32	0\\
33	0\\
34	0\\
35	0\\
36	0\\
37	0\\
38	0\\
39	0\\
40	0\\
41	0\\
42	0\\
43	0\\
44	0\\
45	0\\
46	0\\
47	0\\
48	0\\
49	0\\
50	0\\
51	0\\
52	0\\
53	0.49748274494065\\
54	0.986643937444899\\
55	0.863978464468049\\
56	0.967718196402377\\
57	0.831106260875559\\
58	0.750520595818172\\
59	0.733333803915634\\
60	0.535952820088559\\
61	1.1229036462113\\
62	1.63673630233326\\
63	1.43724654942627\\
64	1.52140102958622\\
65	1.42500013803234\\
66	1.65049054714717\\
67	1.56838322005518\\
68	1.68015827266594\\
69	1.50312443917874\\
70	1.31728872327006\\
71	1.24386182164615\\
72	1.4486828641961\\
73	2.05334731423976\\
74	1.86354800034646\\
75	1.76161820354764\\
76	1.56428333908475\\
77	1.36502574802222\\
78	1.2973232010384\\
79	1.18460594082567\\
80	1.20060416199676\\
81	1.3458467751413\\
82	1.58772271141127\\
83	1.47198665887254\\
84	1.30510803451307\\
85	1.11403650644951\\
86	1.49485630089129\\
87	1.3342554930345\\
88	1.3641513754138\\
89	1.52421652032958\\
90	2.30294240590739\\
91	2.27114070516734\\
92	2.07227442703124\\
93	2.0201333866568\\
94	1.82027374285789\\
95	2.06951877809516\\
96	1.97657839440552\\
97	1.77667956039529\\
98	1.62410202682989\\
99	1.42753366391748\\
100	1.228105978658\\
};
\addlegendentry{Unencrypted}

\addplot [color=mycolor2, line width=1.2pt, dashed]
  table[row sep=crcr]{%
0	0\\
1	0.8\\
2	0.60033124\\
3	0.40615293\\
4	0.20909057\\
5	0.00982498\\
6	0\\
7	0\\
8	0\\
9	0\\
10	0\\
11	0\\
12	0\\
13	0\\
14	0\\
15	0\\
16	0\\
17	0\\
18	0\\
19	0\\
20	0\\
21	0\\
22	0\\
23	0\\
24	0\\
25	0\\
26	0\\
27	0\\
28	0\\
29	0\\
30	0\\
31	0\\
32	0\\
33	0\\
34	0\\
35	0\\
36	0\\
37	0\\
38	0\\
39	0\\
40	0\\
41	0\\
42	0\\
43	0\\
44	0\\
45	0\\
46	0\\
47	0\\
48	0\\
49	0\\
50	0\\
51	0\\
52	0\\
53	0.49705801\\
54	0.9871205\\
55	0.86407126\\
56	0.96668127\\
57	0.83063968\\
58	0.75090992\\
59	0.73495092\\
60	0.53742101\\
61	1.12170837\\
62	1.63827062\\
63	1.43871583\\
64	1.52142072\\
65	1.42587896\\
66	1.6499404\\
67	1.56848289\\
68	1.6810757\\
69	1.5038767\\
70	1.31813306\\
71	1.24451331\\
72	1.44913652\\
73	2.05374552\\
74	1.86386588\\
75	1.76221084\\
76	1.56482205\\
77	1.36560605\\
78	1.29839341\\
79	1.18512366\\
80	1.2020933\\
81	1.34619286\\
82	1.5894825\\
83	1.47317699\\
84	1.30666599\\
85	1.11540824\\
86	1.49498593\\
87	1.33466657\\
88	1.36497058\\
89	1.52473062\\
90	2.30364298\\
91	2.27190702\\
92	2.07303598\\
93	2.02049198\\
94	1.82069647\\
95	2.07699672\\
96	1.97831161\\
97	1.77884985\\
98	1.6190501\\
99	1.42043394\\
100	1.2204359\\
};
\addlegendentry{Encrypted}

\addplot [color=red, dotted, line width=1.2pt]
  table[row sep=crcr]{%
0	0.1\\
1	0.1\\
2	0.1\\
3	0.1\\
4	0.1\\
5	0.1\\
6	0.1\\
7	0.1\\
8	0.1\\
9	0.1\\
10	0.1\\
11	0.1\\
12	0.1\\
13	0.1\\
14	0.1\\
15	0.1\\
16	0.1\\
17	0.1\\
18	0.1\\
19	0.1\\
20	0.1\\
21	0.1\\
22	0.1\\
23	0.1\\
24	0.1\\
25	0.1\\
26	0.1\\
27	0.1\\
28	0.1\\
29	0.1\\
30	0.1\\
31	0.1\\
32	0.1\\
33	0.1\\
34	0.1\\
35	0.1\\
36	0.1\\
37	0.1\\
38	0.1\\
39	0.1\\
40	0.1\\
41	0.1\\
42	0.1\\
43	0.1\\
44	0.1\\
45	0.1\\
46	0.1\\
47	0.1\\
48	0.1\\
49	0.1\\
50	0.1\\
51	0.1\\
52	0.1\\
53	0.1\\
54	0.1\\
55	0.1\\
56	0.1\\
57	0.1\\
58	0.1\\
59	0.1\\
60	0.1\\
61	0.1\\
62	0.1\\
63	0.1\\
64	0.1\\
65	0.1\\
66	0.1\\
67	0.1\\
68	0.1\\
69	0.1\\
70	0.1\\
71	0.1\\
72	0.1\\
73	0.1\\
74	0.1\\
75	0.1\\
76	0.1\\
77	0.1\\
78	0.1\\
79	0.1\\
80	0.1\\
81	0.1\\
82	0.1\\
83	0.1\\
84	0.1\\
85	0.1\\
86	0.1\\
87	0.1\\
88	0.1\\
89	0.1\\
90	0.1\\
91	0.1\\
92	0.1\\
93	0.1\\
94	0.1\\
95	0.1\\
96	0.1\\
97	0.1\\
98	0.1\\
99	0.1\\
100	0.1\\
};
\addlegendentry{Threshold}

\end{axis}
\end{tikzpicture}%